%% file: draft_Zc4020_multi_channel_joint_analysis.tex
\documentclass[twocolumn,superscriptaddress,prx]{revtex4-2}
\usepackage{graphicx}
\usepackage{epstopdf}
\usepackage{units}
\usepackage{multirow}
\usepackage{bigstrut}
\usepackage{overpic}
\usepackage{array}
\usepackage{color}
\usepackage{amsmath}
\usepackage{xspace}
\usepackage{amsfonts}
\usepackage{subfigure}
\usepackage{verbatim}
\usepackage{footmisc}
\usepackage{color}
\usepackage{epstopdf}
\usepackage[colorlinks,
            linkcolor=blue,
            anchorcolor=blue,
            citecolor=blue]{hyperref}
\RequirePackage{lineno}
\usepackage{appendix}
\usepackage{tabularx}
\usepackage{xcolor}

\newcommand{\dst}{D^{*}}
\newcommand{\dstbar}{\bar{D}^{*}}
\newcommand{\dstzero}{D^{*0}}

\newcommand{\dstplus}{D^{*+}}
\newcommand{\dstminus}{D^{*-}}

\newcommand{\dzerobar}{\bar{D}^{0}}
\newcommand{\dzero}{D^{0}}
\newcommand{\ee}{e^+e^-}

\newcommand{\kaonm}{K^-}
\newcommand{\kp}{K^+}
\newcommand{\km}{K^-}

\newcommand{\pip}{\pi^+}
\newcommand{\pim}{\pi^-}
\newcommand{\pizero}{\pi^0}

\newcommand{\jpsi}{J/\psi}

\newcommand{\mev}{\,\unit{MeV}}

\newcommand{\mevcc}{\,\unit{MeV}/c^2}
\newcommand{\gev}{\,\unit{GeV}}
\newcommand{\gevc}{\,\unit{GeV}/c}
\newcommand{\gevcc}{\,\unit{GeV}/c^2}
\newcommand{\invpb}{\,\unit{pb}^{-1}}

\newcommand {\eg}{\textit{e.g.}}
\newcommand {\ie}{\textit{i.e.}}

\newcommand{\zc}{T_{c\bar{c}}}
\newcommand{\zcs}{T_{c\bar{c}\bar{s}}}
\newcommand{\zcone}{T_{c\bar{c}1}}

\newcommand{\zcp}{T_{c\bar{c}}^{+}}
\newcommand{\zcm}{T_{c\bar{c}}^{-}}

\newcommand{\dstdstpi}{D^{*0}D^{*-}\pi^{+}}

\newcommand{\pipijpsi}{\pi^{+}\pi^{-}\jpsi}
\newcommand{\hc}{h_{c}}
\newcommand{\Ri}{R_{i}}

\newcommand{\pipihc}{\pi^{+}\pi^{-}\hc}

\newcommand{\pipi}{\pip\pim}
\newcommand{\kk}{\kp\km}
\newcommand{\ppbar}{p\Bar{p}}
\newcommand{\ks}{K_{S}^{0}}
\newcommand{\etac}{\eta_{c}}
\newcommand{\lp}{l^{+}}
\newcommand{\lm}{l^{-}}

\newcommand{\jp}{J^{P}}
\newcommand{\chisq}{\chi^{2}}
\newcommand{\gammas}{\gamma^{*}}

\usepackage{dcolumn}
\newcolumntype{d}[1]{D{.}{\pm}{#1}}

\usepackage{titlesec}
\titleformat{\subsection}{\normalfont\bfseries\centering}{}{0em}{}
\titleformat{\subsubsection}{\normalfont\bfseries\centering}{}{0em}{}

\begin{document}%%

%\setpagewiselinenumbers
%\modulolinenumbers[2]
%\linenumbers
\title{\boldmath Multi-channel joint analysis of the exotic charmonium-like state $T_{c\bar{c}}(4020)$}

\date{\it \small \bf \today}

\author{
\begin{small}
\begin{center}
\input{authorlist_2025-03-21.tex}
\end{center}
\end{small}
}

\vspace{4cm}

\begin{abstract}
This paper reports the first multi-channel joint analysis to identify the properties of the exotic charmonium-like state $T_{c\bar{c}}(4020)$ via the electron-positron annihilation process $e^{+}e^{-}\to\pi^{+}T_{c\bar{c}}(4020)^{-}+c.c$. A partial wave analysis is performed simultaneously in three decay channels $T_{c\bar{c}}(4020)^{-}\to {D}^{*0}D^{*-}$, $\pi^{-}J/\psi$, and $\pi^{-}h_{c}$, based on data samples taken at $\sqrt{s}=4.395$ and $4.416\,\mathrm{GeV}$ with an integrated luminosity of $1598.9\,\mathrm{pb}^{-1}$ collected with the BESIII detector operating on the BEPCII collider.
For the first time, the spin-parity of the $T_{c\bar{c}}(4020)^{-}$ is determined to be $J^{P}=1^{+}$ with a significance $11.7\sigma$.
Pole positions are extracted on the Riemann sheets with three branch points in the complex energy plane.
Furthermore, the relative branching fractions are obtained as $\mathcal{B}[T_{c\bar{c}}(4020)^{-}\to\pi^{-}J/\psi]/\mathcal{B}[T_{c\bar{c}}(4020)^{-}\to{D}^{*0}D^{*-}]=(3.6\pm0.6\pm1.6)\times10^{-3}$ and $\mathcal{B}[T_{c\bar{c}}(4020)^{-}\to\pi^{-}h_{c}]/\mathcal{B}[T_{c\bar{c}}(4020)^{-}\to{D}^{*0}D^{*-}]=(8.9\pm1.3\pm2.3)\times10^{-2}$, where the first uncertainties are statistical, and the second are systematic.
\end{abstract}

\maketitle
%%%%%%%%%%%%%%%%%%%%%%%%%%%%%%%%%%%%%%%%%%%%%%%%%%%%%%%%%%%%%%%%
%%%%%     Introduction       Part                  %%%%%%%%%%%%%
%%%%%%%%%%%%%%%%%%%%%%%%%%%%%%%%%%%%%%%%%%%%%%%%%%%%%%%%%%%%%%%%

\subsection{I. INTRODUCTION}
Quantum chromodynamics (QCD) allows for the existence of exotic states beyond typical quark-antiquark mesons and three-quark baryons.
Numerous experiments searched for and explored various exotic states to enhance the understanding of QCD, particularly of the color confinement mechanism.
In recent years, charged charmonium-like states and their neutral isospin partners, known as $\zc$ (also $Z_c$) and  $\zcs$ (also $Z_{cs}$) states, have been observed with different processes in a number of experiments~\cite{ParticleDataGroup:2024cfk,Gross:2022hyw,Olsen:2014qna,Husken:2024rdk}. These states are of great interest as candidates for exotic tetraquarks.

The isotriplet charmonium-like states, $\zcone(3900)$ and $\zc(4020)$ (also $\zc(4025)$) are investigated in various processes. The $\zcone(3900)$ is observed in the processes $\ee\to\pi\pi\jpsi$, $\pi\bar{D}\dst$ and $\pi D\dstbar$ by BESIII~\cite{BESIII:2013ris,BESIII:2013qmu,BESIII:2015pqw, BESIII:2015cld,BESIII:2015ntl}, Belle~\cite{Belle:2013yex}, and with CLEO-c data~\cite{Xiao:2013iha}. In particular, its spin-parity is identified to be $J^P=1^+$ by BESIII~\cite{BESIII:2017bua,BESIII:2020oph}.
For the heavier $\zc(4020)$, it is observed by BESIII~\cite{BESIII:2013ouc,BESIII:2013mhi,BESIII:2014gnk,BESIII:2015tix} with the  $\ee\to\pi\pi\hc$ and $\pi\dst\dstbar$ processes.
These $\zc$ states, composed of at least four quarks of $c\bar{c}q\bar{q}$ ($q$=$u$, $d$), have been interpreted through various theoretical frameworks, such as tetraquark configurations~\cite{Voloshin:2013dpa,Maiani:2013nmn,Ali:2011ug}, hadronic molecules~\cite{Guo:2013sya,Cui:2013yva,Cui:2013xla,Guo:2017jvc}, hadro-charmonium~\cite{Dubynskiy:2008mq,Voloshin:2007dx} and kinematic effect~\cite{Chen:2013wca,Wang:2013cya,Li:2013xia,Liu:2013vfa,Swanson:2014tra} but their underlying structures remain unclear.
To identify the nature of these states, a multi-channel joint analysis with different processes is a necessity.

The relative branching fractions (BFs) for different decay channels of the $\zc$ provide crucial information for distinguishing between models of its internal structure. For example, a charmonium-core scenario would favor its decays into hidden-charm final states ($\eg$, $\pi\jpsi$ and $\pi\hc$), suppressing the open-charm decays. Conversely, a significantly larger BF for open-charm decays ($\eg$, $D\dstbar$ and $\dst\dstbar$) would support a hadronic molecular interpretation.

\begin{figure}[hp]
    \includegraphics[width=0.45\textwidth]{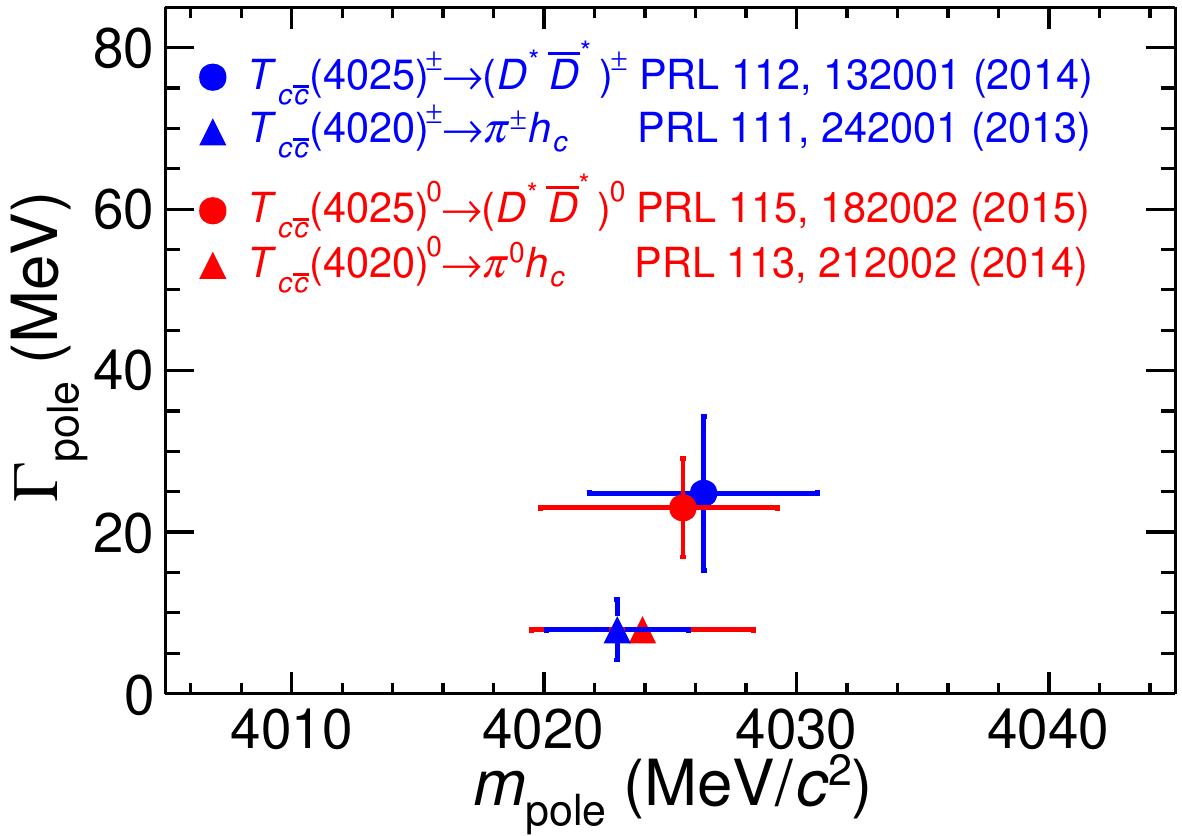}
    \caption{The pole masses and widths of the $\zc(4020)$ (also $\zc(4025)$) states obtained in the processes of $\ee\to\pi\pi\hc$ and $\dst\dstbar\pi$. Here, PRL indicates the reference Physical Review Letters.}
    \label{fig:Zccomparison}
\end{figure}

The mass, width, and spin-parity are fundamental properties of particles, which are highly correlated with their intrinsic nature. The $\zc(4020)$ resonance parameters known so far are obtained through one-dimensional fits of the $\pi\hc$ or $\dst\dstbar$ invariant mass spectra.
However, the discrepancy of measured widths between the two decay channels, as illustrated in Fig.~\ref{fig:Zccomparison}, indicates that the line-shape model used in the one-dimensional mass fit is deficient.
The spin-parity of $\zc$ state is crucial to figure out its nature and can only be determined experimentally. However, unlike $\zcone(3900)$, the spin-parity of $\zc(4020)$ has not been identified in experiment yet.

Simultaneous multi-channel joint partial wave analysis (PWA) provides accurate information on the properties of the hadron state of interest, $\ie$, the resonance parameters, the spin-parity quantum number and the relative BFs among different decay channels.
This paper reports the first multi-channel joint PWA of the $\zc(4020)^-$ combining the three decay channels, $\zc(4020)^{-}\to\dstzero\dstminus$, $\pim\jpsi$, and $\pim\hc$ in the process of $\ee\to\zc(4020)^-\pip$ at the centre-of-mass energies at $\sqrt{s}=4.395$ and $4.416\gev$. Throughout the paper, charge conjugate channels are always implied, unless otherwise specified.
The data sets used are accumulated with the BESIII detector at the BEPCII collider, with an integrated luminosity of $1598.9\invpb$~\cite{BESIII:2022dxl,BESIII:2020eyu}.

%%%%%%%%%%%%%%%%%%%%%%%%%%%%%%%%%%%%%%%%%%%%%%%%%%%%%%%%%%%%%%%%
%%%%%%%%%     Detector       Part                  %%%%%%%%%%%%%
%%%%%%%%%%%%%%%%%%%%%%%%%%%%%%%%%%%%%%%%%%%%%%%%%%%%%%%%%%%%%%%%
\subsection{II. APPARATUS AND DATA SETS }
\subsubsection{A. BEPCII collider and BESIII detector}
The BESIII detector~\cite{BESIII:2009fln} records symmetric $e^+e^-$ collisions provided by the BEPCII storage ring~\cite{Yu:2016cof} in the center-of-mass energy range from 1.85 to $4.95\gev$, with a peak luminosity of $1.1\times10^{33}\,\mathrm{cm}^{-2}\mathrm{s}^{-1}$ achieved at $\sqrt{s} = 3.773\gev$.
BESIII has collected large data samples in this energy region~\cite{BESIII:2020nme}. The cylindrical core of the BESIII detector covers 93\% of the full solid angle and consists of a helium-based multilayer drift chamber (MDC), a plastic scintillator time-of-flight system (TOF), and a CsI(Tl) electromagnetic calorimeter (EMC), which are all enclosed in a superconducting solenoidal magnet providing a $1.0\,\mathrm{T}$ magnetic field.
The solenoid is supported by an octagonal flux-return yoke with resistive plate counter muon identification modules interleaved with steel.
%The acceptance of charged particles and photons is 93\% over $4\pi$ solid angle.
The charged-particle momentum resolution at $1\gevc$ is $0.5\%$, and the
${\rm d}E/{\rm d}x$ resolution is $6\%$ for electrons from Bhabha scattering. The EMC measures photon energies with a resolution of $2.5\%$ ($5\%$) at $1\gev$ in the barrel (end cap) region.
The time resolution in the TOF barrel region is $68\,\mathrm{ps}$, while that in the end cap region is $110\,\mathrm{ps}$.
The end cap TOF system was upgraded in 2015 using multigap resistive plate chamber technology, providing a time resolution of $60\,\mathrm{ps}$~\cite{Li:2017jpg,Guo:2017sjt,Cao:2020ibk}.

\subsubsection{B. Data and Monte Carlo simulation samples}
Three decay channels, $\ee\to\dstdstpi$, $\pipijpsi$, and $\pipihc$, are simultaneously fitted for data taken at $\sqrt{s}=4.395$ and $4.416\gev$.
The candidate events are selected according to the descriptions in the methodology section {\bf A}. The numbers of the survived data ($N_\mathrm{data}$) and background events for different channels are presented in Table~\ref{tab:dataset}. The background is practically negligible in the $\dstzero\dstminus\pip$ channel. For $\pipijpsi$ and $\pipihc$ channels, the backgrounds are estimated based on the event numbers ($N_\mathrm{bkg}$) in the sideband regions of $\jpsi$ and $h_c$ mass spectra, respectively, to be scaled by the area ratio of the signal and sideband regions $\mathcal{W}_\mathrm{bkg}$.
Overall, after subtracting backgrounds, 1430, 641, and 1285 net signal events are fed into multi-channel joint PWA studies for the  $\dstzero\dstminus\pip$, $\pipijpsi$, and $\pipihc$ channels.

\begin{table}[htbp]
\begin{center}
\caption{The number of data events in signal region and background events in sideband region used in the PWA analysis for the process $\ee\to\gamma^*\to X$.}\label{tab:dataset}
\begin{tabular}{ccccccc}\hline\hline
 $\sqrt{s}~(\gev)$ & Process (X) &$\mathcal{L}~(\mathrm{pb}^{-1})$ & $N_\mathrm{data}$ & $N_\mathrm{bkg}$  & $\mathcal{W}_\mathrm{bkg}$\\\hline
\multirow{3}{*}{4.416} &$\dstdstpi$ & \multirow{3}{*}{1090.7} & 1038 & --- & --- \\
&$\pipijpsi$ & & 467 & 227 & 0.25\\
&$\pipihc$ & & 1085 & 821 & 0.5\\\hline
\multirow{3}{*}{4.395} &$\dstdstpi$ & \multirow{3}{*}{508.2} & 392 & --- & --- \\
&$\pipijpsi$ & & 260 & 118 & 0.25\\
&$\pipihc$ & & 495 & 360 & 0.5\\\hline\hline
\end{tabular}
\end{center}
\end{table}

Monte Carlo (MC) simulated samples produced with a {\sc geant4}-based~\cite{GEANT4:2002zbu} software package, which includes the geometric description of the BESIII detector and the detector response, are used to determine the detection efficiencies and to estimate the backgrounds.
The simulation models the beam energy spread and initial-state radiation (ISR) in the $\ee$ annihilation with the generator {\sc kkmc}~\cite{Jadach:1999vf}.
The signal MC samples of the $\ee\to\dstdstpi$, $\pipijpsi$, and $\pipihc$ processes are generated according to a phase-space (PHSP) sampling from $\gammas$ decay. The dedicated decay models are used for specific intermediate particles. The ISR effect in each process obeys the corresponding measured production cross section distribution~\cite{BESIII:2023cmv,BESIII:2016bnd,BESIII:2016adj}.
Possible background contributions are estimated by an inclusive MC simulation sample, which includes the production of open-charm processes, the ISR production of vector charmonium(-like) states, and the continuum processes incorporated in {\sc kkmc}.
All particle decays are modelled with {\sc evtgen}~\cite{Ping:2008zz} using BFs taken from the Particle Data Group~(PDG)~\cite{ParticleDataGroup:2024cfk}, when available, and unknown charmonium decays are modelled with {\sc lundcharm}~\cite{Yang:2014vra}.
Final-state radiation from charged final-state particles is incorporated using {\sc photos}~\cite{Richter-Was:1992hxq}.

%%%%%%%%%%%%%%%%%%%%%%%%%%%%%%%%%%%%%%%%%%%%%%%%%%%%%%%%%%%%%%%%
%%%%%%%%%%%%%%%%%     PWA    Part             %%%%%%%%%%%%%%%%%%
%%%%%%%%%%%%%%%%%%%%%%%%%%%%%%%%%%%%%%%%%%%%%%%%%%%%%%%%%%%%%%%%
\subsection{III. PARTIAL WAVE ANALYSIS}
\subsubsection{A. Amplitude construction}
The PWA used for this study is based upon the generic software package {\sc tf-pwa}~\cite{Jiang:2024vbw}, where the helicity amplitude is represented in the $LS$-coupling scheme in which  the decay amplitudes are given in terms of definite orbital angular momentum $L$ and total intrinsic spin $S$.
For a two-body decay
\begin{figure}[tp]
\begin{center}
    \includegraphics[width=0.78\linewidth]{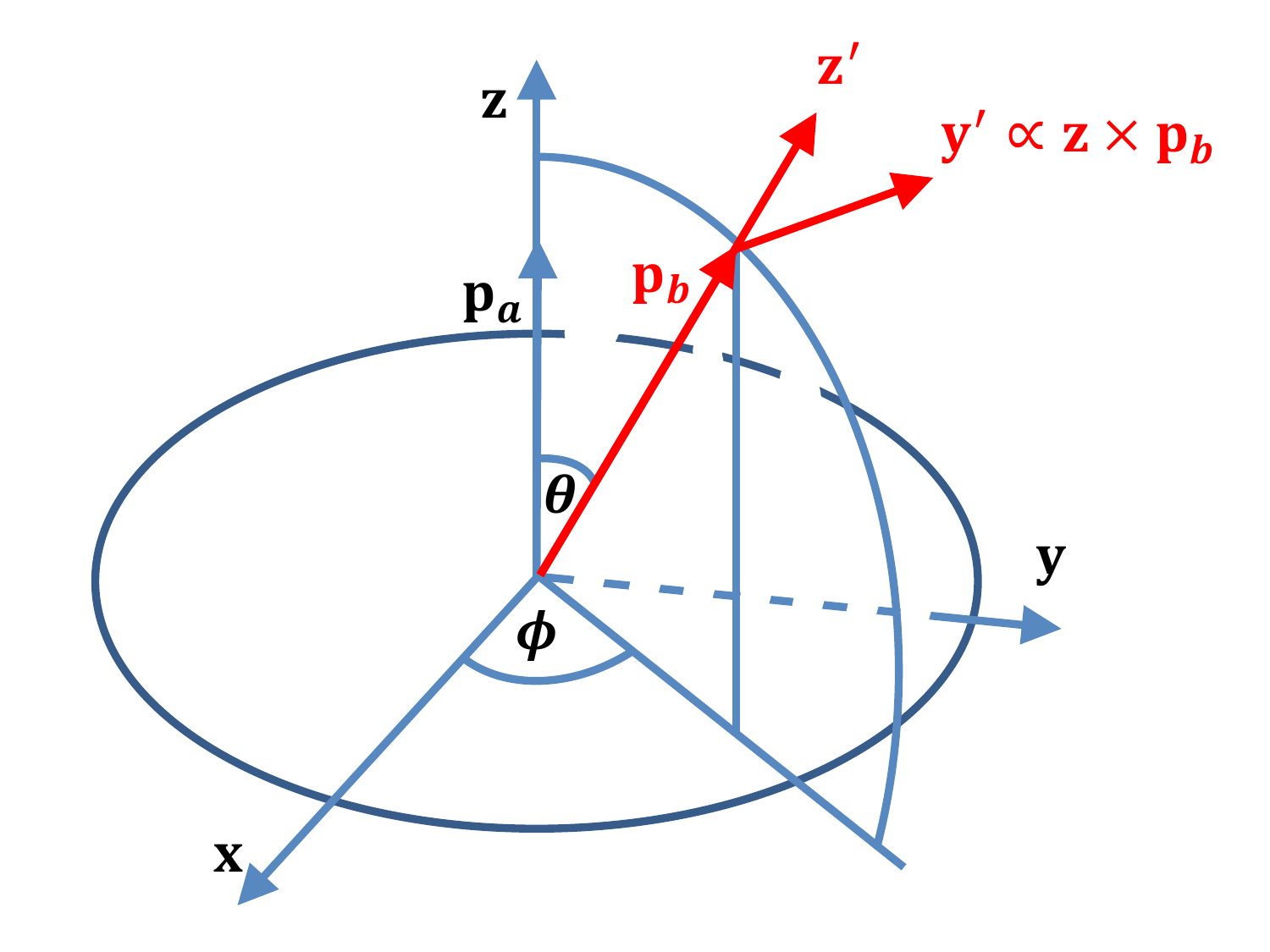}
    \caption{An illustration and definition of the helicity angles involved in the PWA.}
    \label{fig:helicity_angle}
\end{center}
\end{figure}
\begin{equation}
a(J_{a},P_{a})\to b(J_{b},P_{b})+c(J_{c},P_{c}),\nonumber
\end{equation}
with $J$ and $P$ denoting the spin and parity quantum numbers,  its decay amplitude is given by
\begin{equation}
A_{\lambda_{b},\lambda_{c}}=H_{\lambda_{b},\lambda_{c}}D^{J_{a}*}_{\lambda_{a},\lambda_{b}-\lambda_{c}}(\phi,\theta,0),
\end{equation}
where $H_{\lambda_{b},\lambda_{c}}$ is the helicity amplitude and $D$ represents the Wigner $D$-matrix.
The expansion of $H_{\lambda_{b},\lambda_{c}}$ in terms of $LS$-amplitudes conserves parity if (and only if) appropriate combinations of $L$ and $S$ are chosen~\cite{Chung:1997jn,Chung:1993da,Chung:2007nn},
\begin{equation}
\label{chung_forma}
H_{\lambda_{b},\lambda_{c}}=\sum_{LS}C\sqrt{2L+1\over 2J_{a}+1}g_{LS}p^{L}B'_{L}(p,p_{0},d),
\end{equation}
where $C=\langle L0 S\delta|J_{a}\delta\rangle\langle J_{b}\lambda_{b}J_{c}-\lambda_{c}|S\delta\rangle$ is the Clebsch-Gordan coefficient; $\lambda_{b}$ and $\lambda_{c}$ are the helicities of the decay products $b$ and $c$, respectively;
$\delta=\lambda_{b}-\lambda_{c}$; $g_{LS}$ is a complex coupling constant determined by the fit to the data events; $B'_{L}(p,p_{0},d)$ is the reduced Blatt-Weisskopf barrier factor with the centrifugal barrier value $d=3.0\,(\gevc)^{-1}$~\cite{Blatt:1952ije,VonHippel:1972fg}; $p_{(0)}=|{\bf p_{(0)}}|$ is the modulus of the relative momentum between the two daughter particles in their parent rest frame.
In the Wigner $D$-matrix, $\lambda_{a}$ is the helicity of the mother particle $a$. As illustrated in Fig.~\ref{fig:helicity_angle}, the momentum vector $\mathbf{p}_a$ of the mother particle $a$ is aligned with the $\mathbf{z}$-axis of the $\mathbf{xyz}$ system. The flight direction of particle $b$ is taken along the $\mathbf{z'}$-axis of the daughter helicity system, with its $\mathbf{y'}$-axis defined as $\mathbf{y'} = \mathbf{z}\times\mathbf{p}_b/|\mathbf{p}_b|$ and $\mathbf{x'} =\mathbf{y'}\times\mathbf{z'}$, forming a right-handed system $\mathbf{x'y'z'}$. The helicity angles $\theta$ and $\phi$ are defined by rotation to align the $\mathbf{xyz}$ system with the $\mathbf{x'y'z'}$ system.  The detailed formalism of the helicity amplitude is provided in the methodology section {\bf B}.

The intermediate states that couple to the $\pip\pim$ final state with the same spin and parity ($0^+$) are described as $(\pipi)_\mathrm{S-wave}$, which parametrizes the contribution from the wide resonances $\sigma$, $f_0(980)$, and $f_0(1370)$ for $\pip\pim\jpsi$ case. Similarly, for the $\pip\pim\hc$ case, the $(\pipi)_\mathrm{S-wave}$ describes the contributions from $\sigma$ and $f_0(980)$.
The $f_0(1370)$ and $f_2(1270)$ states are described with a relativistic Breit-Wigner propagator. Their masses and widths are fixed to the measured values~\cite{BES:2004twe}.
The E791 propagator~\cite{E791:2000vek} is adopted for $\sigma$ resonance, with a running width
$\Gamma(m)=\sqrt{1-(4m_\pi^2/m^2)}\,\Gamma_0$,
with $\Gamma_{0}=324\mev$ and $\sigma$ mass $m_{0}=478\mevcc$. The variable $m$ refers to the reconstructed two-pion invariant mass obtained for each event.
The $f_0(980)$ is parametrized with the Flatt\'{e}-like formula~\cite{BES:2004mws} taking
into account its main decay channels, $\pip\pim$ and $\kp\km$,
\begin{equation}
\label{flatte}
R(m)=\frac{1}{m^2-m_0^2+i[g_1\rho_{\pip\pim}(m)+g_2\rho_{\kp\km}(m)]},
\end{equation}
where $\rho(m)=2|{\bf p}|/m$, $|{\bf p}|$ is the modulus of the momentum of the $\pip$ or $\kp$ in the $f_0(980)$ rest frame, and $g_1$ and $g_2$ are the coupling strengths, which, together with the mass, are fixed to the BESII measurements~\cite{BES:2004mws}, $\ie$, $m_0=970\mevcc$, $g_1=0.138\gev^{2}$, and $g_2=4.45g_{1}$.

A similar Flatt\'{e}-formula is employed to parametrise the propagator of $\zcone(3900)^{\pm}$ by assuming its decay mode is dominated by $\pi\jpsi$ and $\dst\bar{D}$ channels.
The corresponding parameters are fixed to the BESIII measurements~\cite{BESIII:2017bua}, $\ie$, $m_0=3901.5\mevcc$, $g_1=0.075\gev^{2}$, and $g_2=27.1\cdot g_{1}$.

Considering its decays into $\dstzero\dstminus$, $\pim\jpsi$, and $\pim\hc$ final states~\cite{Gong:2016hlt},  the $\zc(4020)^{-}$ lineshape is parametrized by the Breit-Wigner propagator
\begin{align}\label{eq:zc4020}
R(m)=\frac{1}{m^2-m_0^2+im\,\Gamma(m)},
\end{align}
with the running width
\begin{align}
\Gamma(m)=\Gamma_{\dstzero\dstminus}(m)&+\Gamma_{\pim\jpsi}(m) \\
&+\Gamma_{\pim\hc}(m)+\Gamma_{\rm unknown}(m),\nonumber
\end{align}
where $m_0$ is the resonance mass, $\Gamma_{\dstzero\dstminus}$, $\Gamma_{\pim\jpsi}$, and $\Gamma_{\pim\hc}$ denote the partial decay widths of $\zc(4020)^{-}$ coupling to $\dstzero\dstminus$, $\pim\jpsi$, $\pim\hc$ final states, respectively, and $\Gamma_{\rm unknown}$ represents the remaining width of missing decay channels.
The partial decay widths are calculated by the standard formula for two-body decays:
\begin{align}
\Gamma_{X_{i}}(m)={1\over 8\pi}\overline{\sum_{\lambda_1,\lambda_2}}\left|H^{T_{c\bar{c}}^-\to X_{i}}_{\lambda_1,\lambda_2}\right|^2{|{\bf p}|\over m^{2}},
\end{align}
where $X_i$ denotes different decay channels.
The $|{\bf p}|$ is the magnitude of final state momentum,
and $H^{T_{c\bar{c}}^-\to~X_{i}}_{\lambda_1,\lambda_2}$ represents the helicity amplitude of the $i$-th two-body decay.
The symbol $\overline{\sum}_{\lambda_1,\lambda_2}$ indicates the sum over $\lambda_1$, $\lambda_2$, and the average runs over $J$, the spin of $T_{c\bar{c}}^{-}$.
If $m$ is below the mass threshold of the final state, the momentum magnitude ${\bf |p|}$ is replaced by ${i{\bf |p|}}$, where $i$ is the imaginary unit, to preserve the continuity in the complex plane~\cite{Gong:2016hlt}.
To consider the contribution from unknown missing decays, the sum of the helicity amplitude, $\overline{\sum}|H_{X_{\rm unknown}}|^2$, is assumed as a free constant determined by fitting, and the momentum dependence is taken as that used in the $\pim\jpsi$ mode to compute the relative momentum for the missing channels. The charge-conjugated state $\zc(4020)^+$ is treated with the same contribution, with all of the parameters shared with $\zc(4020)^-$.
The non-resonance (NR) propagator is taken as a constant without any dynamics.

The fit fraction (FF) of a given intermediate process is estimated with the truth-level PHSP MC events as
\begin{equation}
\mathrm{FF}_{i} = \frac{\sum_{n\in \rm{PHSP}}|A_i(x_{n})|^{2}}{\sum_{n\in \rm{PHSP}}|\sum_{k}A_k(x_{n})|^{2}},
\end{equation}
where $A_i$ is the amplitude of the $i$-th subprocess, and the integration is carried out with the MC sampling method.
The fraction of the interference between the $i$-th and $j$-th components is calculated by
\begin{equation}
\mathrm{FF}_{i,j}=\frac{\sum_{n\in \rm{PHSP}}|A_i(x_{n})+A_j(x_{n})|^{2}}{\sum_{n\in \rm{PHSP}}|\sum_{k}A_k(x_{n})|^{2}}-\mathrm{FF}_{i}-\mathrm{FF}_{j}.
\end{equation}
The statistical uncertainty of FF is estimated according to the covariance matrix elements $V_{ij}$ from the fit as
\begin{equation}
\delta Y^{2} = \sum_{i=1}^{N_\mathrm{pars}}\sum_{j=1}^{N_\mathrm{pars}}\left({\partial Y\over \partial \boldsymbol{X}_{\it i}}{\partial Y\over \partial \boldsymbol{X}_{\it j}}\right)_{\boldsymbol{X}={\boldsymbol{\mu}}}\cdot V_{ij}({\boldsymbol X}),
\end{equation}
where ${\boldsymbol X}$ is the vector of the fit parameters, and ${\boldsymbol{\mu}}$ contains the fitted values for all parameters. The sum runs over all $N_\mathrm{pars}$ parameters.

\subsubsection{B. Amplitude fitting}
The amplitude fit is based on the maximum likelihood method, whose detailed construction is provided in the methodology section {\bf C}. To provide a unified description of the data samples for the three signal processes, the peak around 4.02 $\gevcc$ in the $M(\dstzero\dstminus)$, $M(\pim\jpsi)$, and $M(\pim h_c)$ invariant-mass spectra is attributed to the $\zc(4020)^-$ resonance, with a spin-parity assignment of $1^+$. We demonstrate that data support this hypothesis, and the possibilities of other assignments are tested in Section \hyperref[sec:jptest]{IV, B}.

For the $\ee\to\dstdstpi$ process, possible components are $\zc(4020)^-\pip$ and $\mathrm{NR}^{(0^{-},\,1^{-},\,2^{-},\,2^{+})}\dstzero$.  Here, $\mathrm{NR}^{(J^P)}$ denotes a non-resonant contribution in $\dstminus\pip$ with spin-parity $J^P$; the same model applied to $\dstzero\pip$ yields the consistent results given the identical spin-parity. Although the $D_{1}(2420)\dst$  production threshold $4.432\gev$ is above the center-of-mass energies of the data, but considering the intrinsic width of the $D_1(2420)$, it is still considered at $\sqrt{s}=4.416\gev$.
For the $\ee\to\pipijpsi$ process, there are $\zc(4020)^\pm\pi^\mp$, $\zcone(3900)^\pm\pi^\mp$, $(\pip\pim)_\mathrm{S-wave}\jpsi$, and $f_2(1270)\jpsi$.
For the $\ee\to\pipihc$ process, $\zc(4020)^\pm\pi^\mp$, $\zcone(3900)^\pm\pi^\mp$, and $(\pip\pim)_\mathrm{S-wave}\hc$ are included.

\begin{table*}[htbp]
\caption{Multi-channel BW parameters of $\zc(4020)^{-}$ from the nominal PWA fit.
The complex coupling strength for $(L,S)$ partial wave is defined as $g_{LS}=\rho_{LS}e^{i\theta_{LS}}$ ($\theta_{LS}$  in radians).
The fitted resonance mass is $m_{0}=4029.4\pm2.8~\mevcc$.}
\label{tab:S1_gls_result}
\begin{center}
\begin{tabular}{c r@{\,=\,}r@{.}l@{\,$\pm$\,}r@{.}l r@{\,=\,}r@{.}l@{\,$\pm$\,}r@{.}l r@{\,=\,}r@{.}l@{\,$\pm$\,}r@{.}l c}
	\hline\hline
	$\rho_{LS},\theta_{LS}$ & \multicolumn{5}{c}{$\zc(4020)^-\to\dstzero\dstminus$} & \multicolumn{5}{c}{$\to\pim\jpsi$} & \multicolumn{5}{c}{$\to\pim\hc$} & $\to~X$ \\\hline
	$\rho_{LS}$  & $\rho_{01}$&11&6&2&6   & $\rho_{01}$&0&3&0&1 & $\rho_{11}$&\,4&6&1&0  &$\overline{\sum}|H_{X_{\rm{unknown}}}|^2=9.5\pm3.7$\\
	$\theta_{LS}$  & $\theta_{01}$&\multicolumn{4}{l}{0\,(\rm{fixed})}  & $\theta_{01}$&\multicolumn{4}{l}{0\,(\rm{fixed})}  & $\theta_{11}$&\multicolumn{4}{l}{0\,(\rm{fixed})} &\\
	$\rho_{LS}$    & $\rho_{21}$&7&0&7&2       & $\rho_{21}$&0&2&0&1      &\multicolumn{5}{c}{ } & \\
	$\theta_{LS}$  & $\theta_{21}$&3&6&1&5     & $\theta_{21}$&5&6&0&4    &\multicolumn{5}{c}{ } & \\
	$\rho_{LS}$    & $\rho_{22}$&33&9&12&8     &\multicolumn{5}{c}{ }     &\multicolumn{5}{c}{ } & \\
	$\theta_{LS}$  & $\theta_{22}$&4&5&2&1     &\multicolumn{5}{c}{ }     &\multicolumn{5}{c}{ } & \\\hline\hline
\end{tabular}
\end{center}
\end{table*}

\begin{figure*}[htbp]
  \begin{minipage}[t]{0.9\textwidth}
    \centering
    \includegraphics[width=0.245\linewidth]{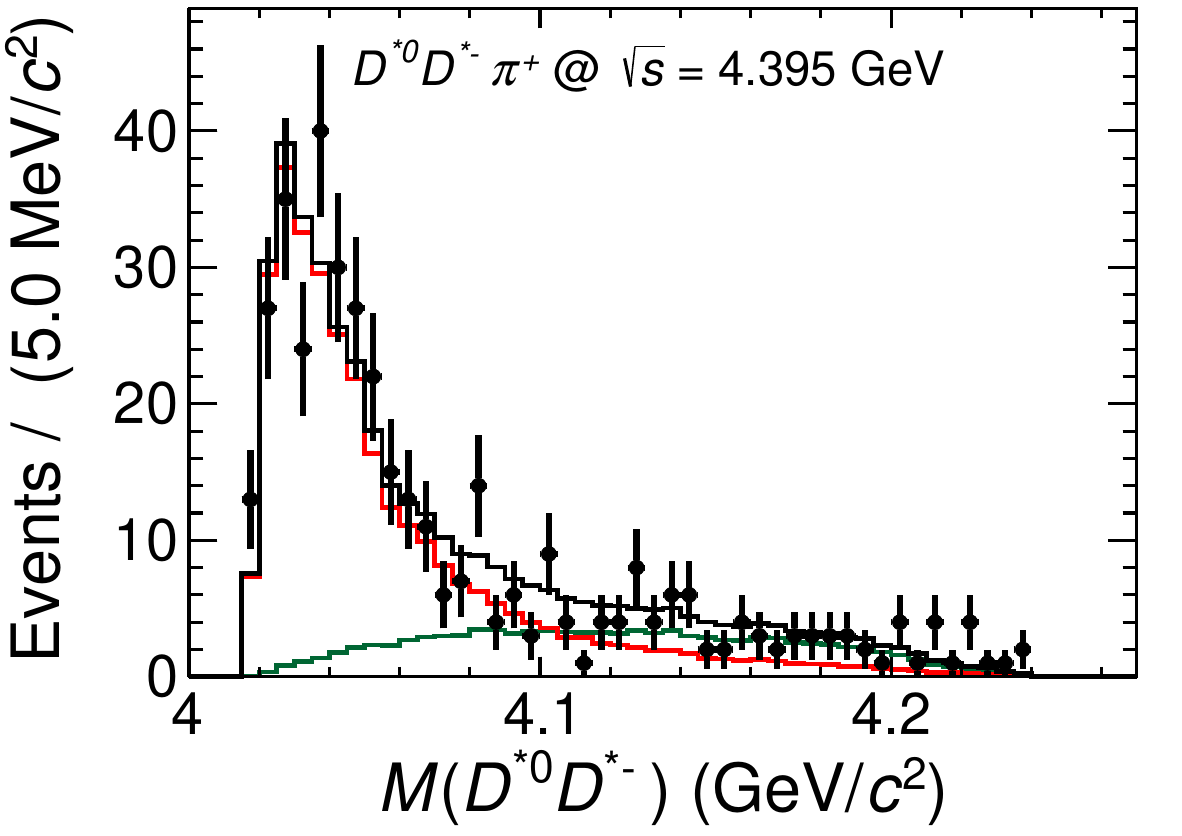}
    \includegraphics[width=0.245\linewidth]{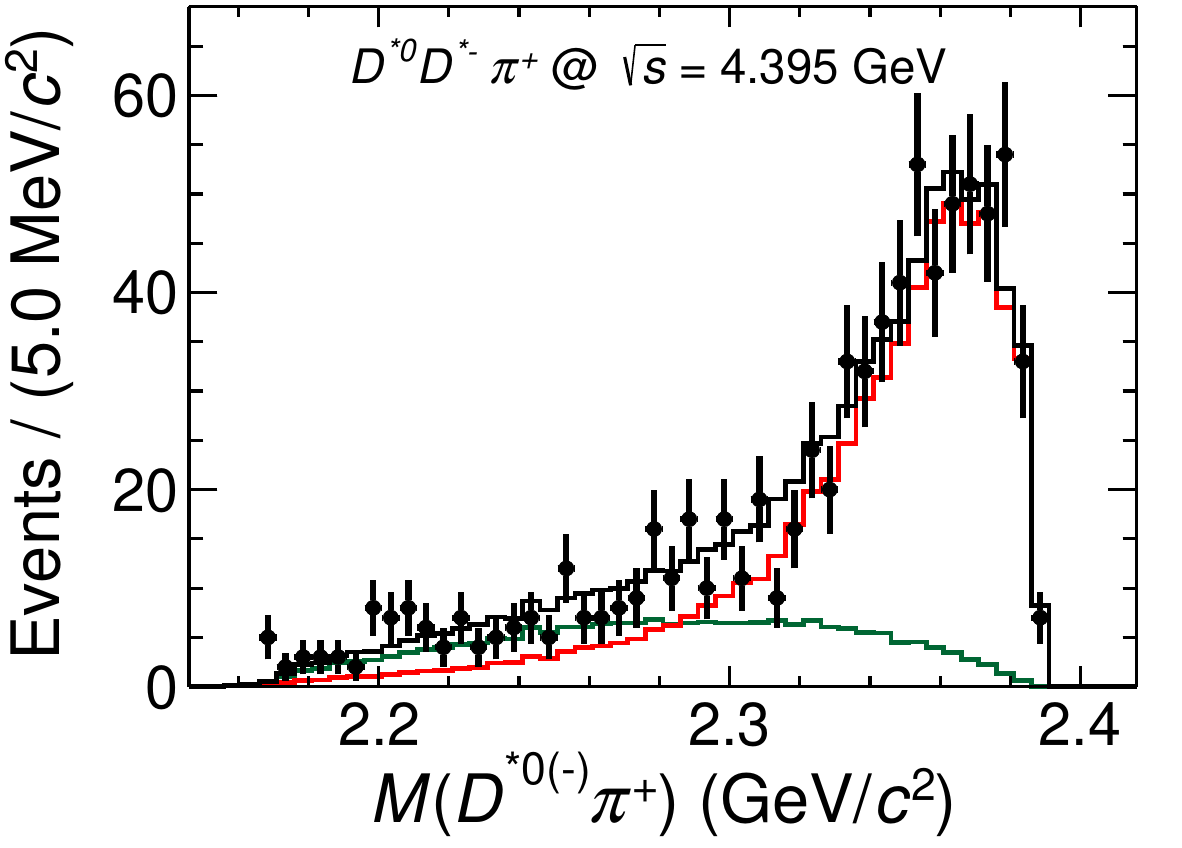}
    \includegraphics[width=0.245\linewidth]{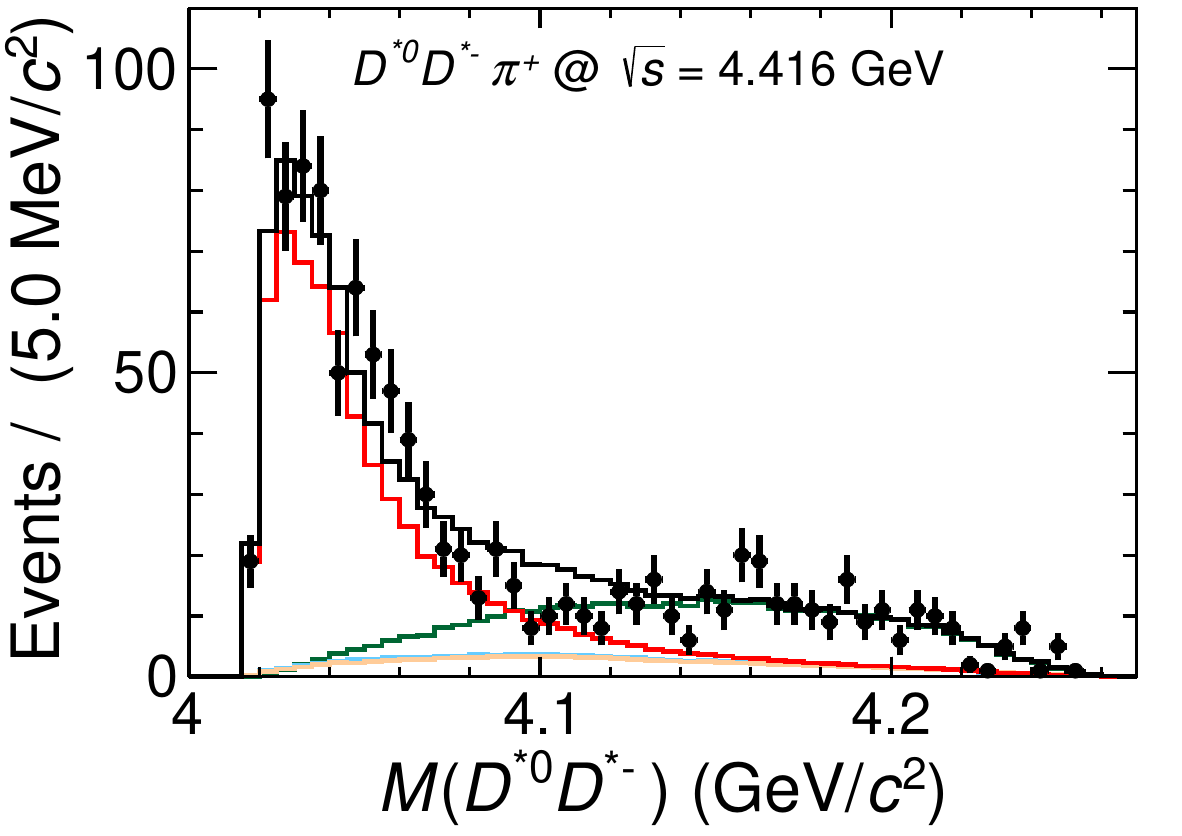}
    \includegraphics[width=0.245\linewidth]{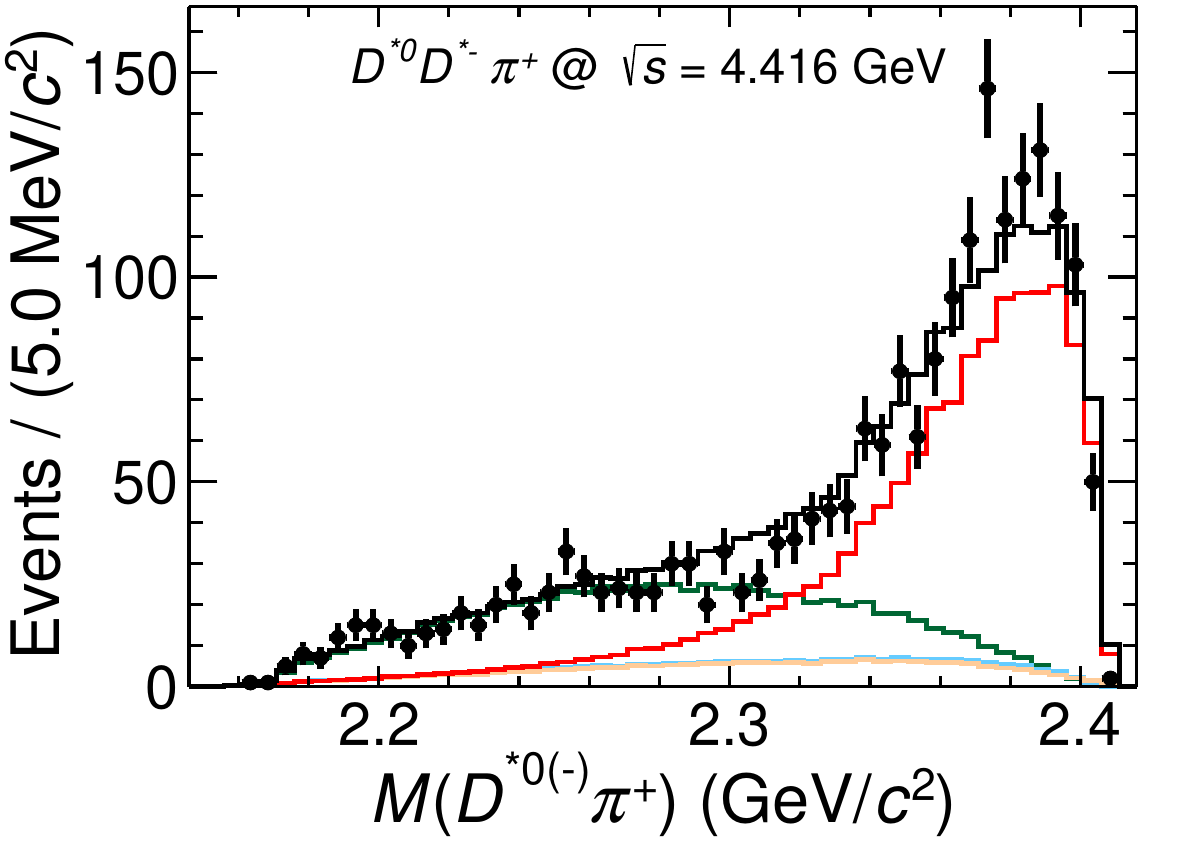}\\
    \includegraphics[width=0.245\linewidth]{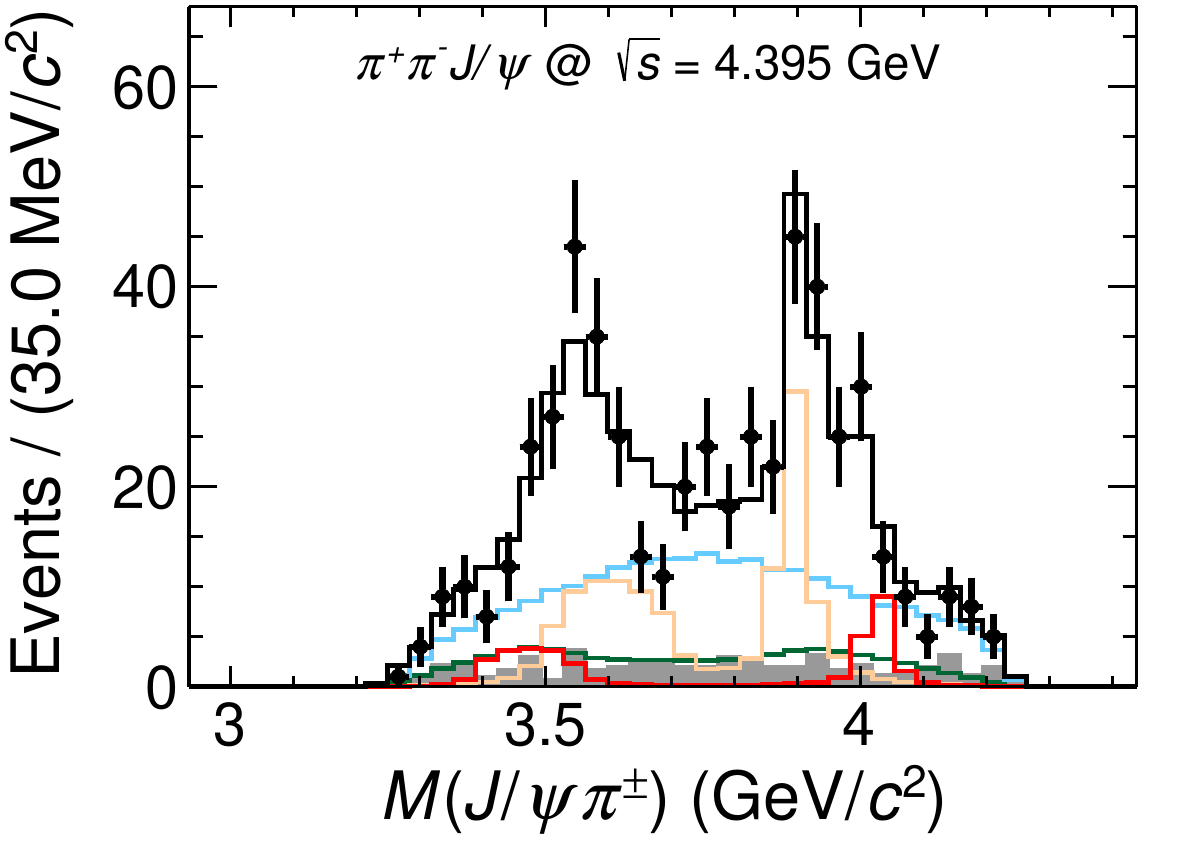}
    \includegraphics[width=0.245\linewidth]{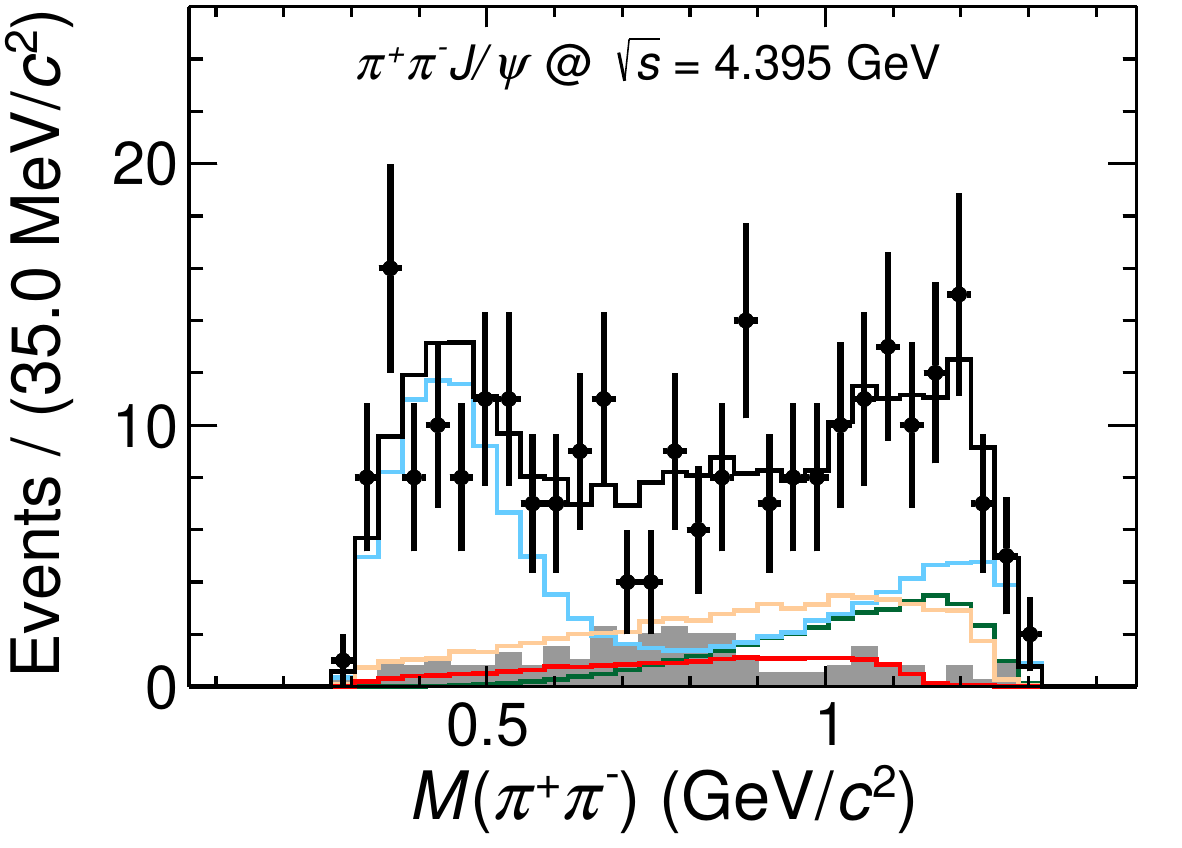}
    \includegraphics[width=0.245\linewidth]{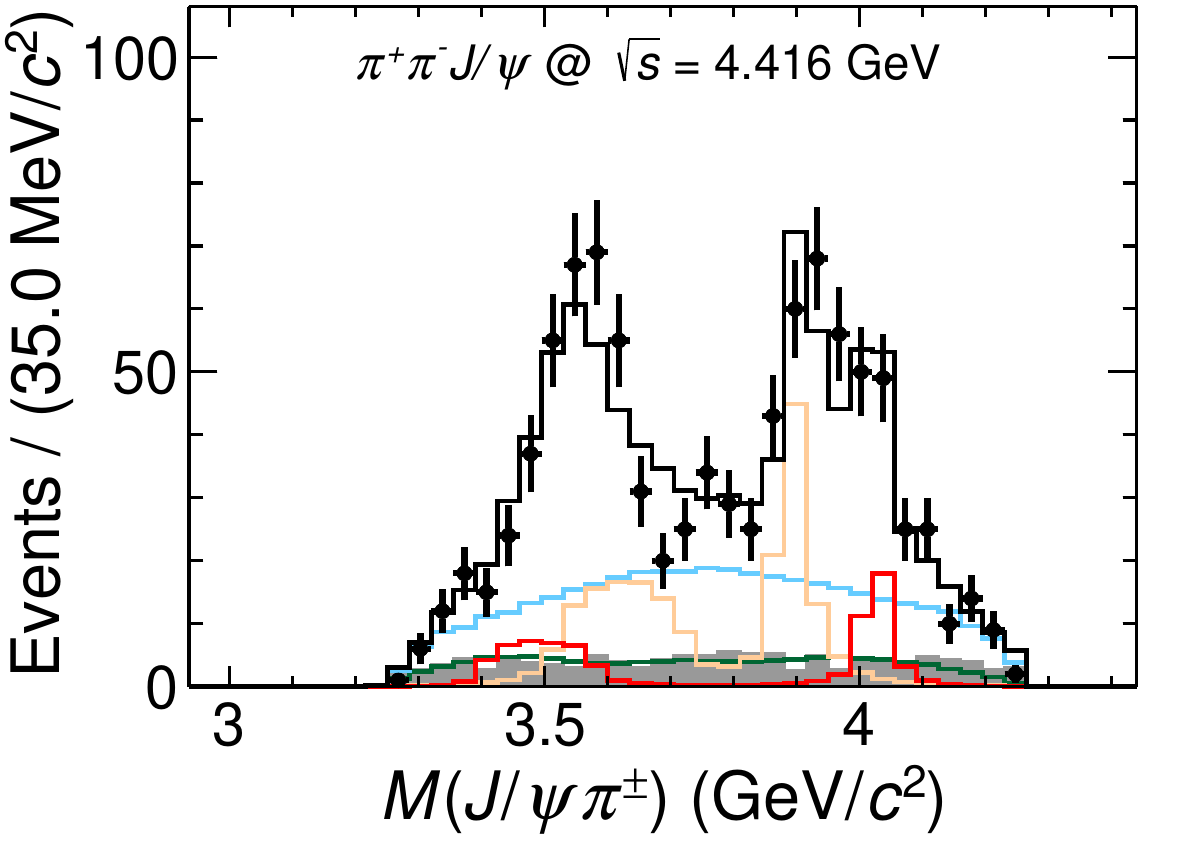}
    \includegraphics[width=0.245\linewidth]{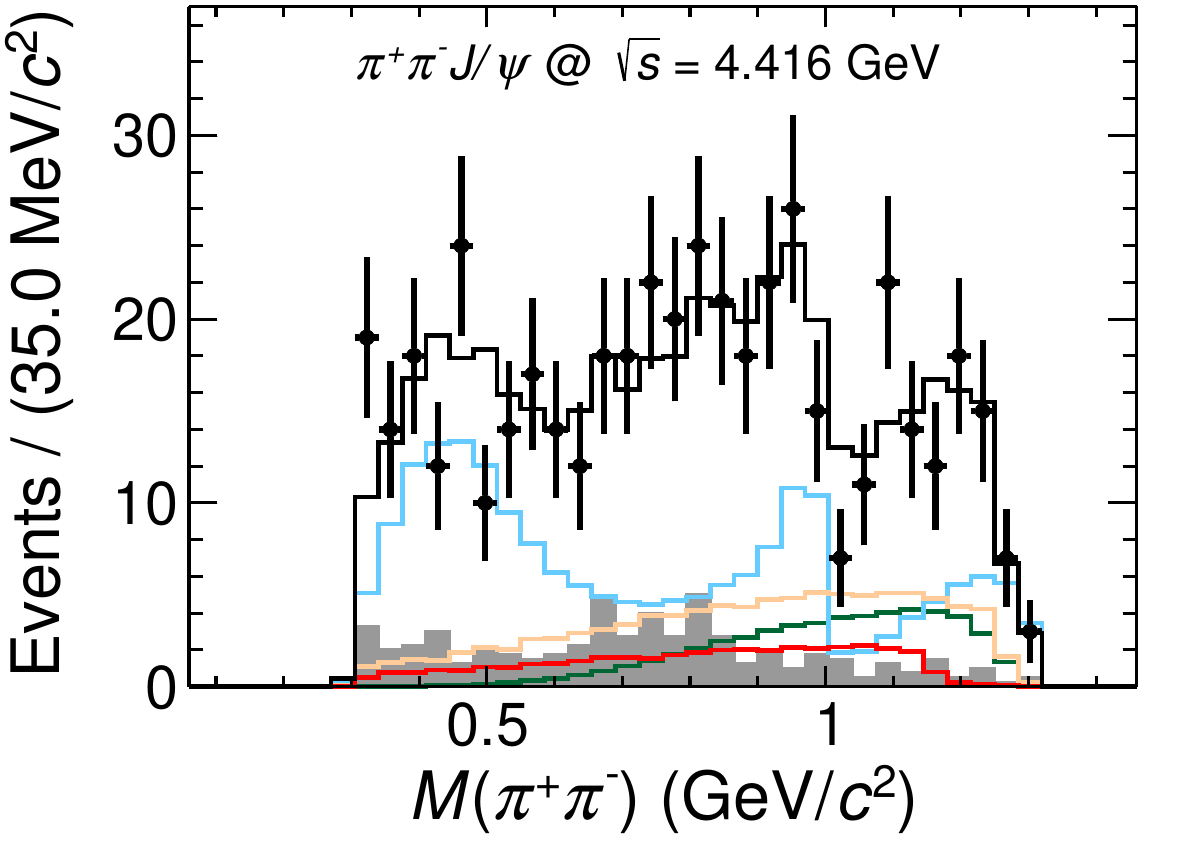}\\
    \includegraphics[width=0.245\linewidth]{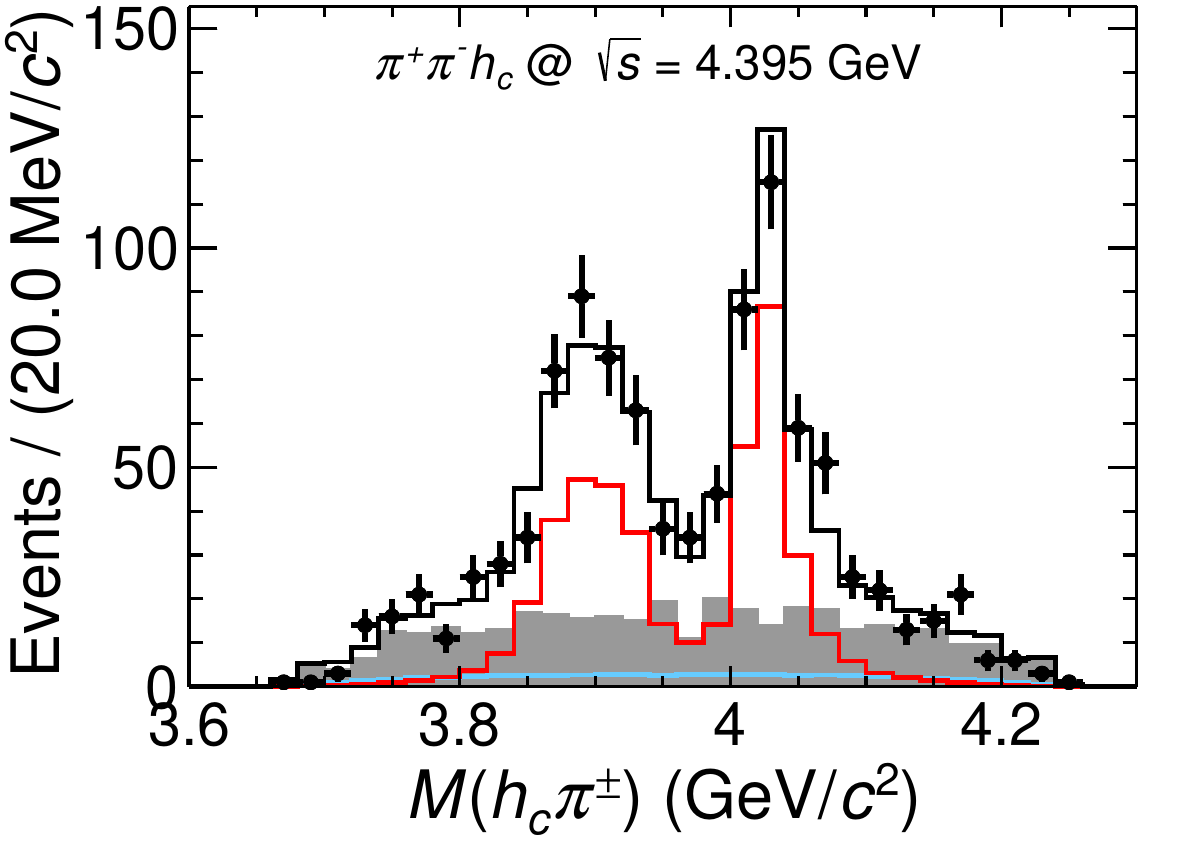}
    \includegraphics[width=0.245\linewidth]{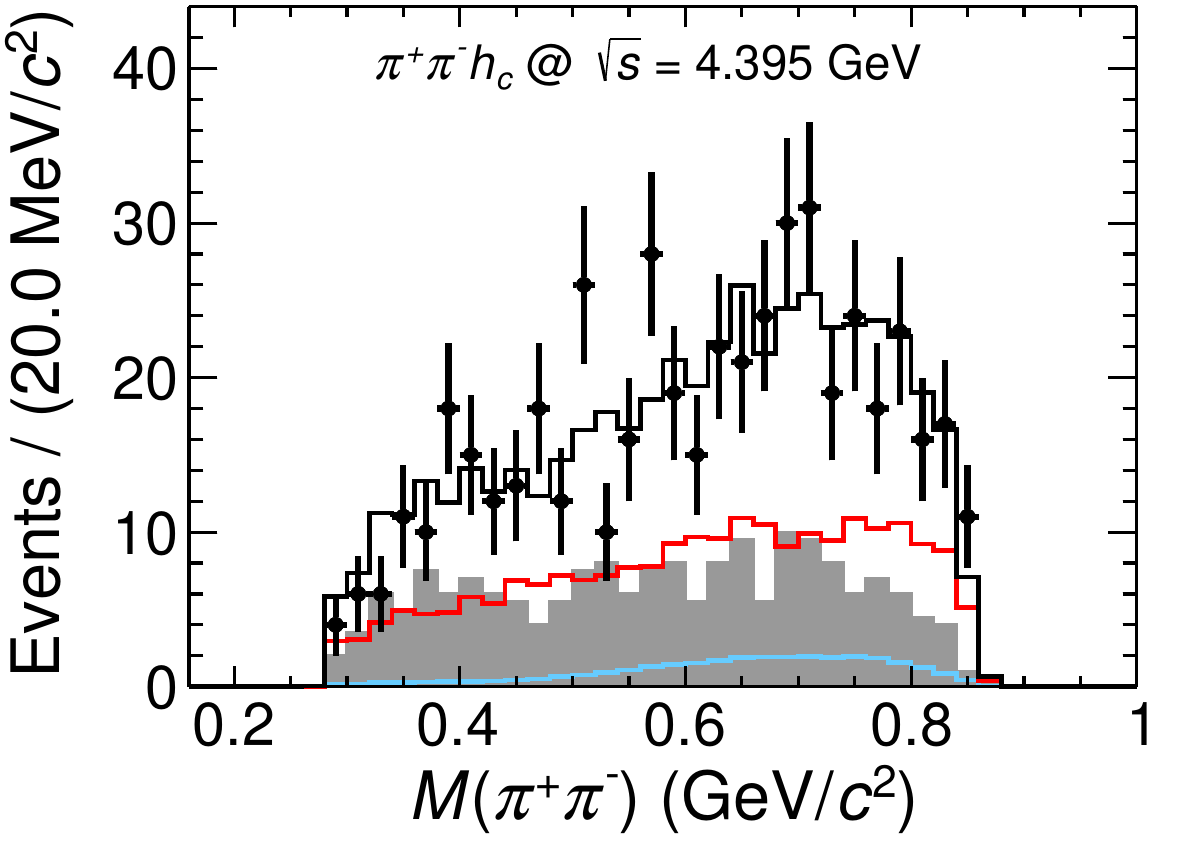}
    \includegraphics[width=0.245\linewidth]{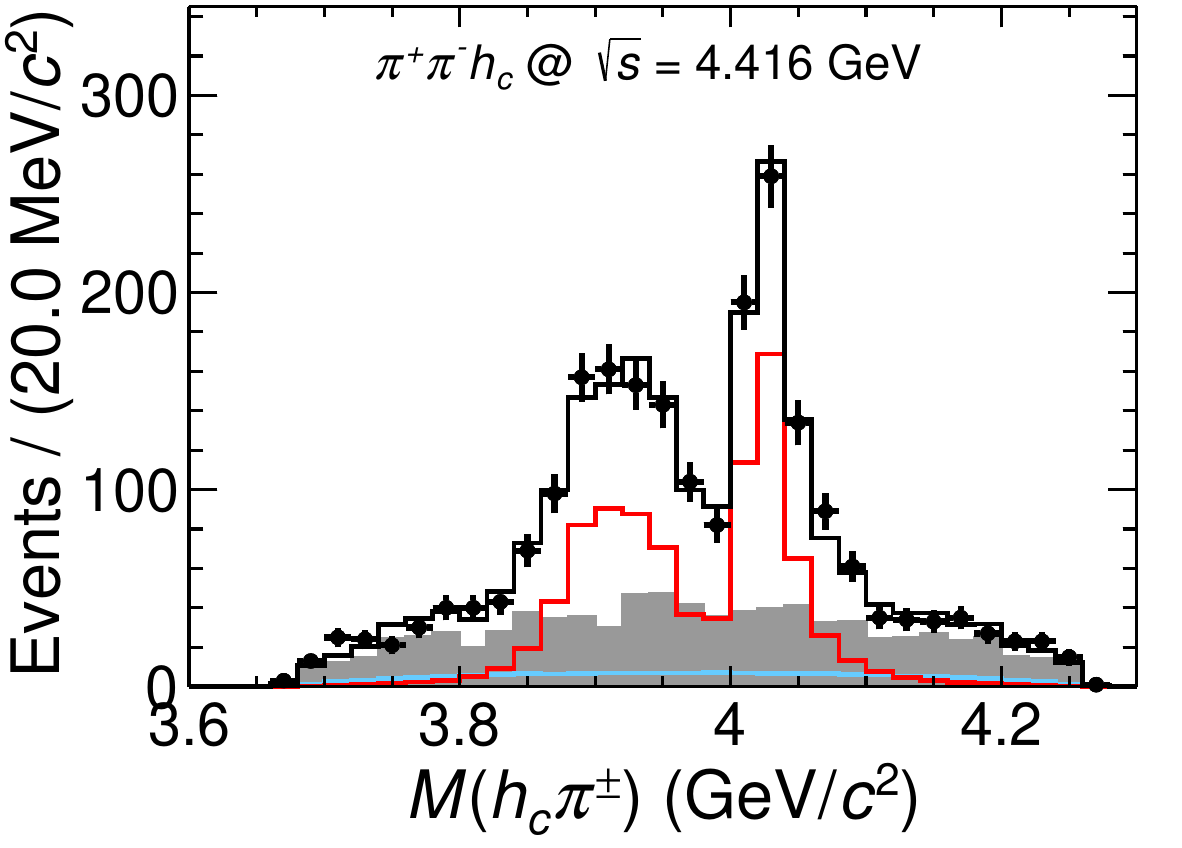}
    \includegraphics[width=0.245\linewidth]{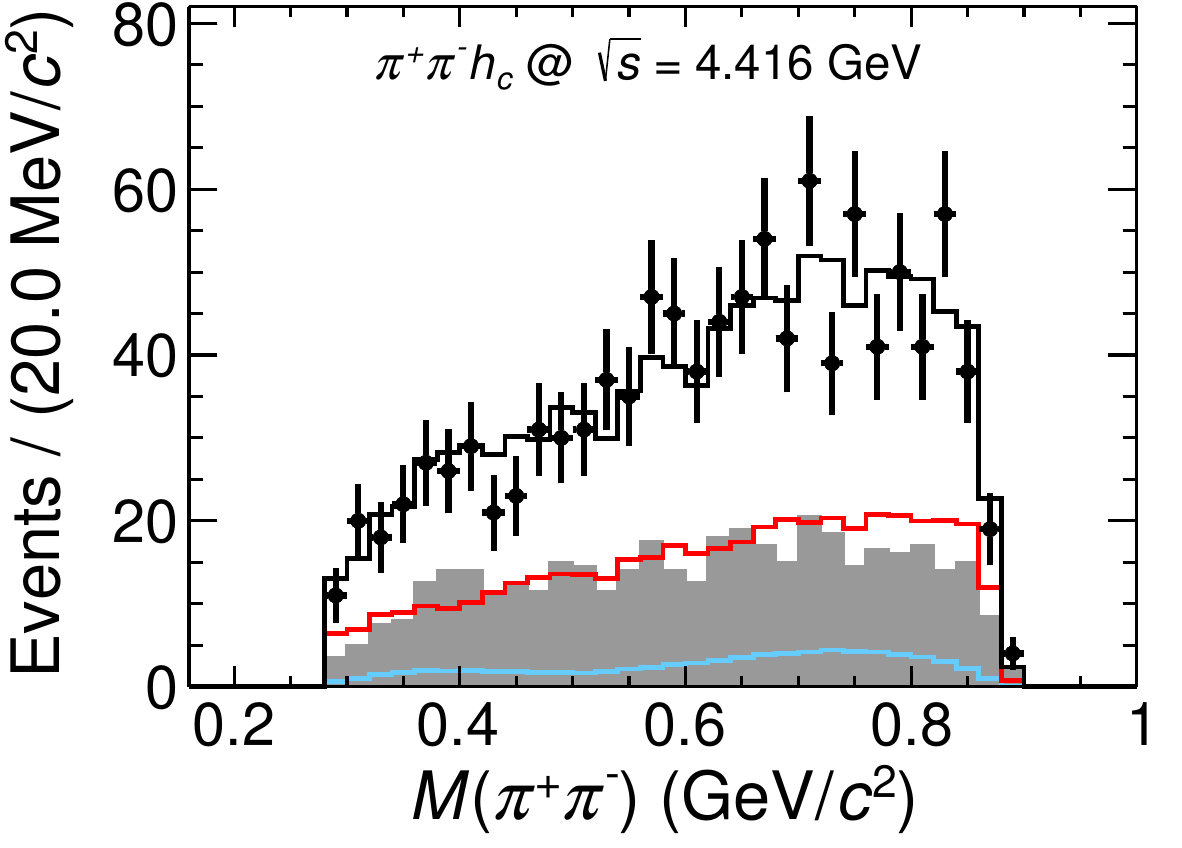}\\
  \end{minipage}% remove space
  \hfill
  \begin{minipage}[t]{0.1\textwidth}
    \centering
    \includegraphics[width=1.8cm, height=2.8cm]{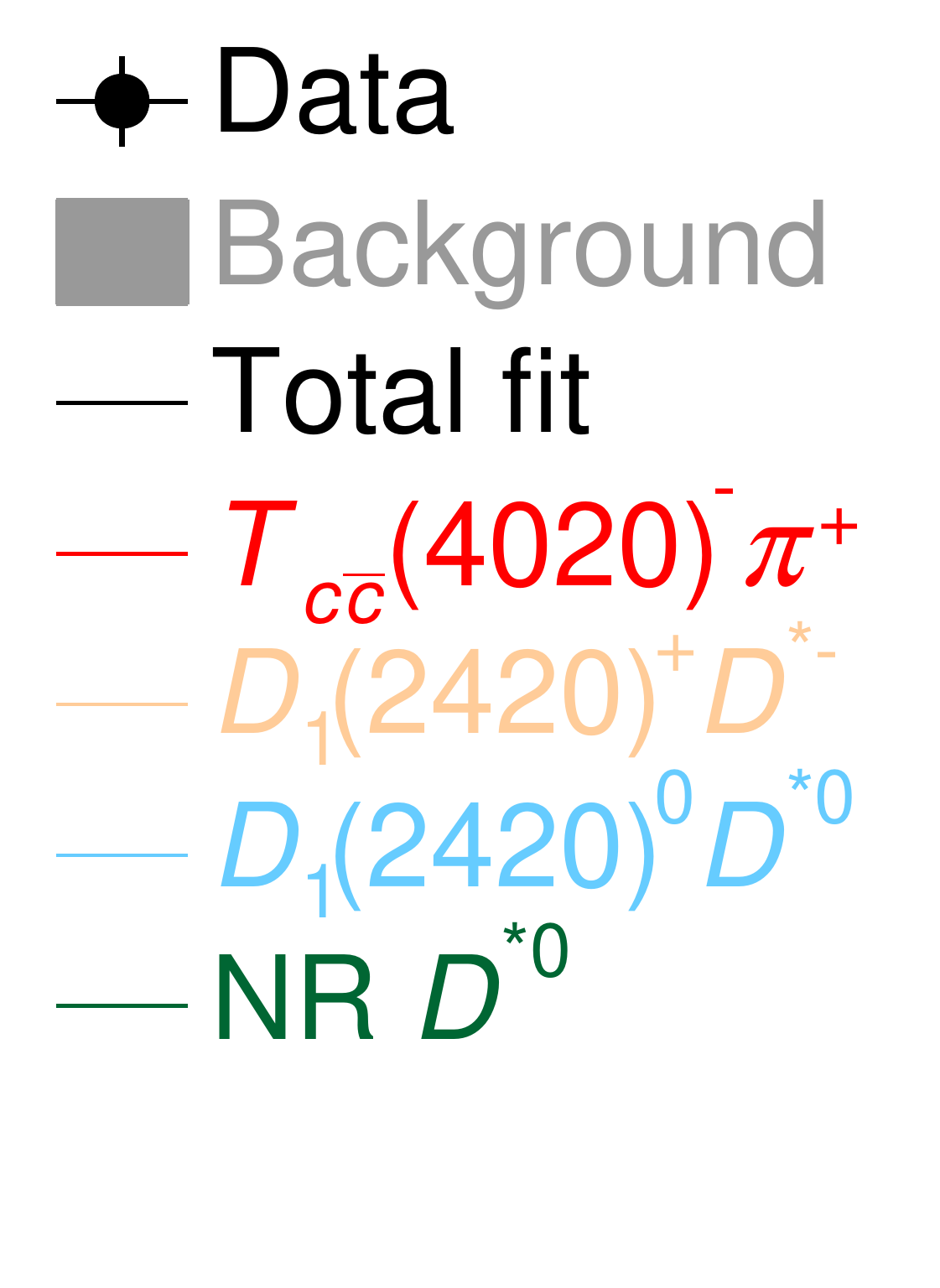}\\
    \includegraphics[width=1.8cm, height=2.8cm]{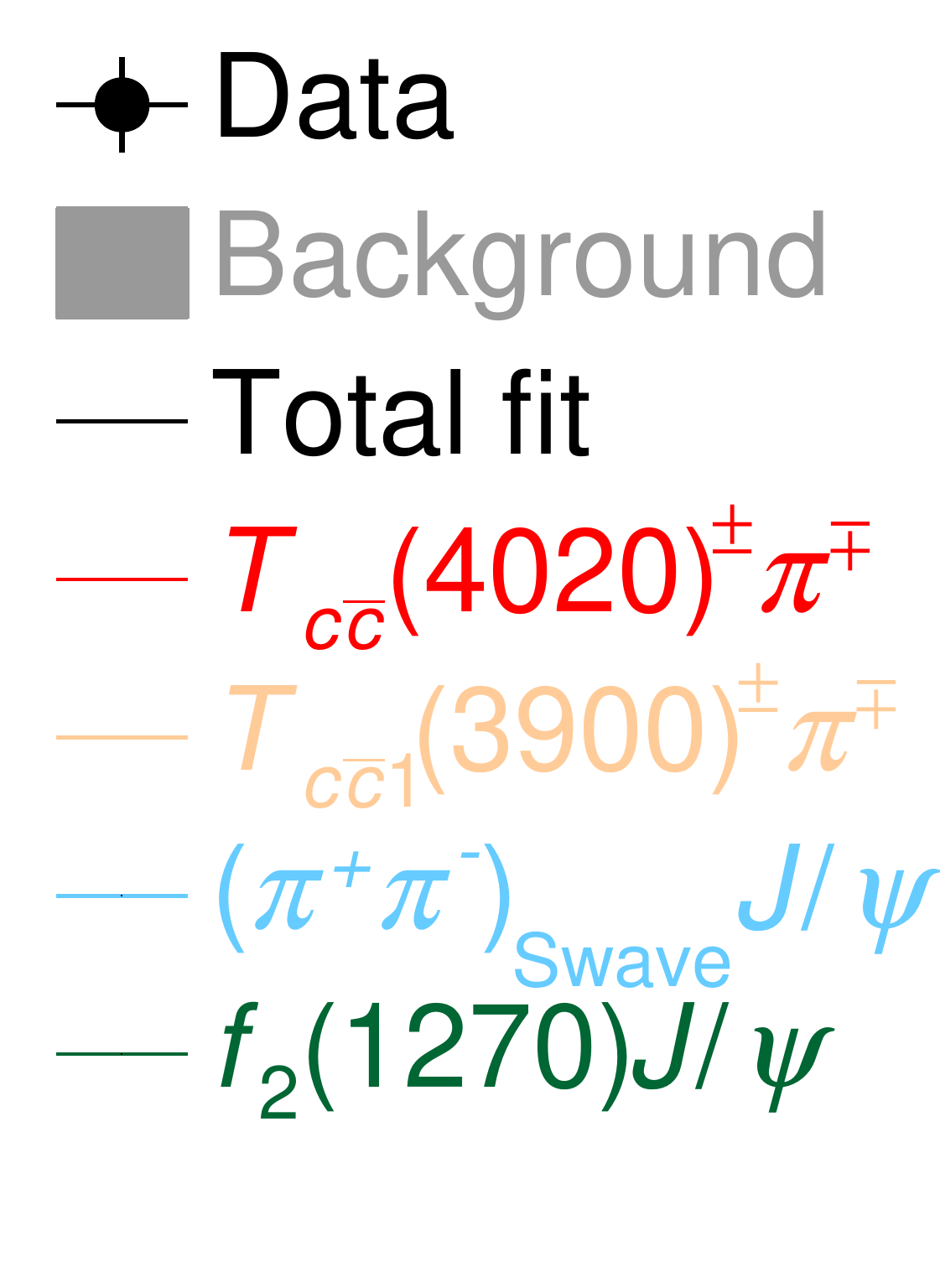}\\
    \includegraphics[width=1.8cm, height=2.8cm]{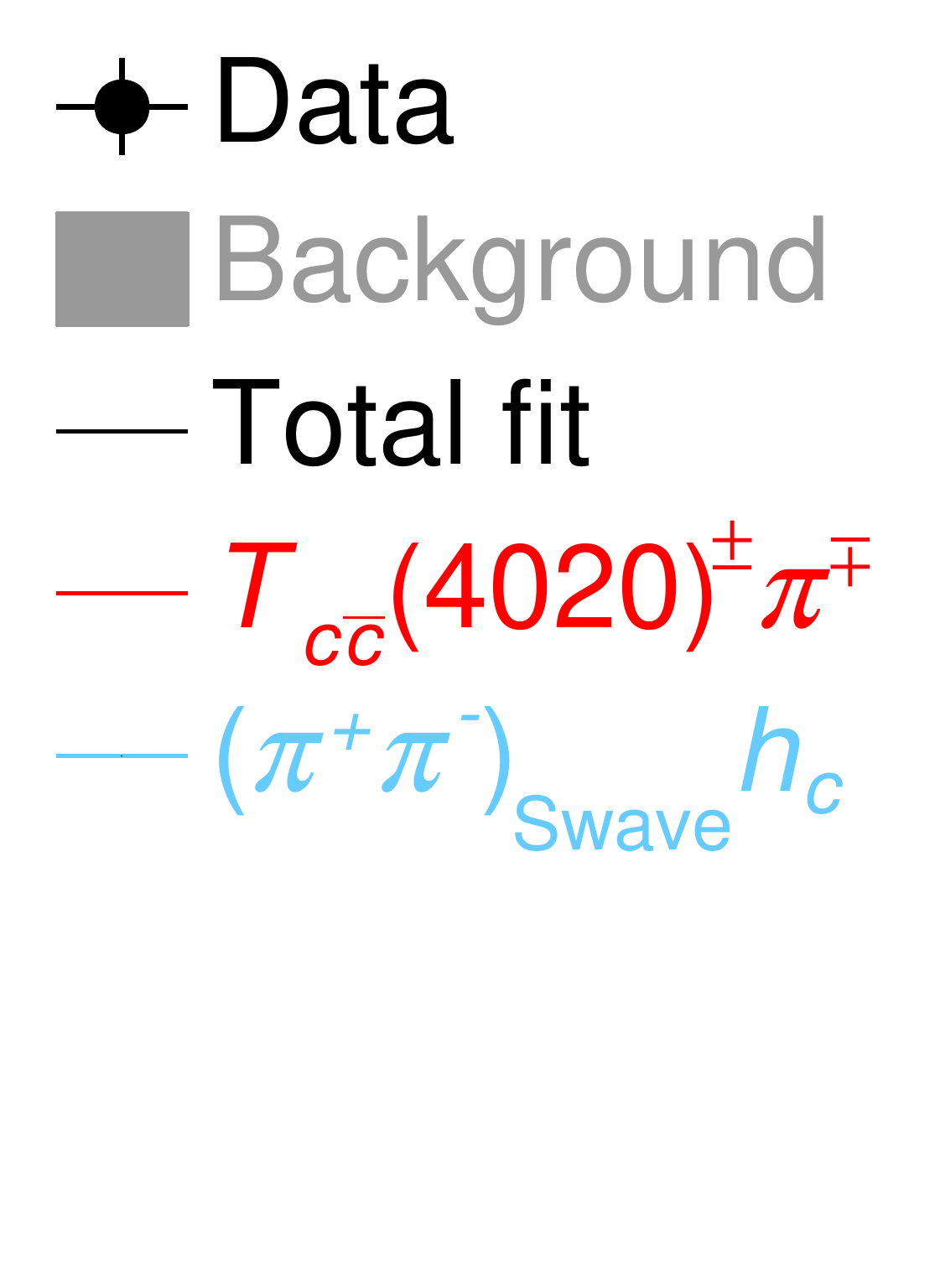}\\
  \end{minipage}
    \caption{Mass projections $M(\dstzero\dstminus)$, $M(D^{*0(-)}\pip)$, $M(\jpsi\pi^{\pm})$, $M(\hc\pi^{\pm})$, and $M(\pipi)$ of the data compared to the results of the simultaneous fit for the processes of $\ee\to\dstdstpi$, $\pipijpsi$, and $\pipihc$ at $\sqrt{s}=4.395$ and $4.416\gev$. The points with error bars are data, and the black histograms are the total fit results, including backgrounds. The grey-shaded histograms denote backgrounds. The colorful lines correspond to the individual contributions from each component.}
    \label{fig:S1_mass_4420_4400}
\end{figure*}

To establish the baseline solution, we employ an iterative method. The initial PWA fit incorporates all potential physical components. The statistical significance of each component is assessed by the difference between negative log-likelihood values of including or excluding a particular resonance in the fit to the data, $\Delta(-2\ln \mathcal{L})$, considering the change in the number of degrees of freedom, $\Delta(\rm{ndof})$.  In each subsequent iteration, the least significant component is removed if its statistical significance falls below 3$\sigma$. Besides, any remaining NR component is added to the fit if its significance exceeds 5$\sigma$. The iteration process ends when all retained components satisfy the above criteria. The final baseline solution includes $\zc(4020)^{\pm}\pi^{\mp}$ and $\zcone(3900)^{\pm}\pi^{\mp}$ resonances, the non-resonance $\mathrm{NR}^{(0^{-},\,1^{-},\,2^{-})}\dstzero$, $(\pip\pim)_\mathrm{S-wave}\jpsi$, $f_2(1270)\jpsi$, and $(\pip\pim)_\mathrm{S-wave}\hc$.
To verify the robustness of the solution, the significance of each component excluded from the baseline is retested, and none is found to meet the significance criteria. These components will be considered for the assessment of systematic uncertainties related to the fitting model.

In the simultaneous fit performed at the two energy points
$\sqrt{s}=4.395$ and $4.416\gev$, the coupling parameters that dominate the production of the $\zc(4020)^-$ state are constrained separately for each point. In contrast, the common parameters governing its decay are shared between them.

\subsection{IV. RESULTS AND DISCUSSIONS}
\subsubsection{A. Pole positions and branching fractions}
The extracted parameters of the fit related to the $\zc(4020)^-$ are reported in Table~\ref{tab:S1_gls_result}. With the fitted $m_{0}$, $\overline{\sum}|H_{X_{\rm{unknown}}}|^2$, and $g_{LS}$ parameter values, the $\zc(4020)^-$ pole positions are extracted in the $\sqrt{s}$-complex plane by constraining its inverse of Breit-Wigner function to zero, $\ie$, $1/R(m_{\rm pole})=0$.
The physically reasonable pole positions in the form of $m_\mathrm{pole} - i\Gamma_\mathrm{pole}/2$ are:
\begin{eqnarray}
m_{\rm pole1}&=&4022.44\pm1.55\mevcc\nonumber,\\
\Gamma_{\rm pole1}&=&38.54\pm2.94\mev\nonumber,
\end{eqnarray}
and
\begin{eqnarray}
\label{S1:pole}
m_{\rm pole2}&=&4023.01\pm1.35\mevcc\nonumber,\\
\Gamma_{\rm pole2}&=&35.02\pm2.20\mev.\nonumber
\end{eqnarray}
A complete listing of all mathematical solutions is shown in Table~\ref{tab:S1_pole} in the methodology section \textbf{D}.

Figure~\ref{fig:S1_mass_4420_4400} shows the projections of the PWA fit results in different mass spectra and their comparisons with data at the two energy points. Numerical results are listed in Table~\ref{tab:S1_FF_4400} and Table~\ref{tab:S1_FF_4420}; the statistical uncertainties are calculated from the covariance matrix of the fit. The product of the Born cross section and the BF for $\zc(4020)^-\to X_i$ is obtained as
\begin{eqnarray}\label{eq:cal_xsBr}
\sigma^B(&&\ee\to\zc(4020)^{-}\pip+c.c.)\cdot \mathcal{B}[\zc(4020)^{-}\to X_{i}]\nonumber\\
&&=\sigma^B(\ee\to\pip X_{i}+c.c.)\cdot\mathrm{FF}_{X_i},
\end{eqnarray}
where $X_{i}$ stands for the mode $\dstzero\dstminus$, $\pim\jpsi$, or $\pim\hc$, while $\mathrm{FF}_{X_i}$ represents the FFs of $\zc(4020)^{-}\to\dstzero\dstminus$, $\zc(4020)^{\pm}\to \pi^{\pm}\jpsi$, and $\zc(4020)^{\pm}\to\pi^{\pm}\hc$, respectively.
The results are listed in Table~\ref{tab:summary_xs}, in which the total three-body Born cross sections are taken from Refs.~\cite{BESIII:2023cmv,BESIII:2016adj,BESIII:2016bnd}.

\begin{table*}[htbp]
    \caption{The measured FFs (\%) for each component for the signal process at $\sqrt{s}=4.395\gev$. Off-diagonal values are the interference FFs.}\label{tab:S1_FF_4400}
    \begin{center}
\begin{tabular}{l   r@{.}l@{\,$\pm$\,}r@{.}l    r@{.}l@{\,$\pm$\,}r@{.}l    r@{.}l@{\,$\pm$\,}r@{.}l     r@{.}l@{\,$\pm$\,}r@{.}l     r@{.}l@{\,$\pm$\,}r@{.}l    r@{.}l@{\,$\pm$\,}r@{.}l}
    \hline\hline
    $\dstdstpi$ & \multicolumn{4}{c}{$\zc(4020)^{-}\pip$}  &  \multicolumn{4}{c}{$\mathrm{NR}\,D^{*0}$} \\ \hline
    \hspace{0.2cm}     $\zc(4020)^{-}\pip$    & 75&1&4&3   &  \multicolumn{2}{c}{ }-    \\
    \hspace{0.2cm}     $\mathrm{NR}\,D^{*0}$  & $-$3&9&3&0 & 28&8&5&9    \\
    \hline\hline
    $\pipi\jpsi$ & \multicolumn{4}{c}{$\zc(4020)^{-}\pi^{+}$} & \multicolumn{4}{c}{$\zc(3900)^{-}\pi^{+}$} & \multicolumn{4}{c}{$\zc(4020)^{+}\pi^{-}$} & \multicolumn{4}{c}{$\zc(3900)^{+}\pi^{-}$} & \multicolumn{4}{c}{$(\pi^{+}\pi^{-})_\mathrm{S-wave}J/\psi$} &  \multicolumn{4}{c}{$f_{2}J/\psi$} \\\hline
    \hspace{0.2cm} $\zc(4020)^{-}\pi^{+}$  & 4&4&2&4     & \multicolumn{2}{c}{ }- & \multicolumn{4}{c}{ }- & \multicolumn{4}{c}{ }- & \multicolumn{4}{c}{ }- &  \multicolumn{4}{c}{ }\hspace{6.9em}-\\
    \hspace{0.2cm} $\zc(3900)^{-}\pi^{+}$  & $-$3&6&1&0      &  12&8&3&1    & \multicolumn{2}{c}{ }- & \multicolumn{4}{c}{ }- & \multicolumn{4}{c}{ }- &  \multicolumn{4}{c}{ }\hspace{6.9em}-\\
    \hspace{0.2cm} $\zc(4020)^{+}\pi^{-}$  & 0&2&0&1         &   0&2&0&2    &  4&4&\multicolumn{2}{l}{\hspace{-0.25em}1.4}  & \multicolumn{2}{c}{ }-  & \multicolumn{4}{c}{ }- &  \multicolumn{4}{c}{ }\hspace{6.9em}-\\
    \hspace{0.2cm} $\zc(3900)^{+}\pi^{-}$  & 0&2&0&2         &   0&7&0&5    &  $-$3&6&\multicolumn{2}{l}{\hspace{-0.25em}1.0}  & 12&8&3&1  & \multicolumn{2}{c}{ }- &  \multicolumn{4}{c}{ }\hspace{6.9em}-\\
    \hspace{0.2cm} $(\pi^{+}\pi^{-})_\mathrm{S-wave}J/\psi$  & $-$1&1&1&7   &  $-$0&3&2&8  &  $-$1&2&\multicolumn{2}{l}{\hspace{-0.25em}1.7}  &  $-$0&3&2&8  &  60&0&21&5  &  \multicolumn{4}{c}{\hspace{0.55em}-}\\
    \hspace{0.2cm} $f_{2}J/\psi$           &  1&2&0&6        &  $-$1&4&1&4  &  1&2&\multicolumn{2}{l}{\hspace{-0.25em}0.6}  &  $-$1&4&1&4  &  0&0&0&0  &  \multicolumn{4}{c}{$15.1\pm5.7$} \\
    \hline\hline
    $\pipi\hc$ &  \multicolumn{4}{c}{$\zc(4020)^{-}\pi^{+}$ } &  \multicolumn{4}{c}{$\zc(4020)^{+}\pi^{-}$} &  \multicolumn{4}{c}{$(\pi^{+}\pi^{-})_\mathrm{S-wave}h_{c}$} \\ \hline
    \hspace{0.2cm}     $\zc(4020)^{-}\pi^{+}$  & 35&0&6&4  & \multicolumn{2}{c}{ }-         &  \multicolumn{4}{c}{ }- \\
    \hspace{0.2cm}     $\zc(4020)^{+}\pi^{-}$  &  0&3&0&1  & 35&0&6&4  &  \multicolumn{2}{c}{ }- \\
    \hspace{0.2cm}     $(\pi^{+}\pi^{-})_\mathrm{S-wave}h_{c}$  & 9&7&4&5  &  9&7&4&5  &  10&1&11&4 \\
    \hline \hline
    \end{tabular}
\end{center}
\end{table*}

\begin{table*}[htbp]
    \caption{The measured FFs (\%) for each component for the signal process at $\sqrt{s}=4.416\gev$. Off-diagonal values are the interference FFs.}\label{tab:S1_FF_4420}
    \begin{center}
    \begin{tabular}{l   r@{.}l@{\,$\pm$\,}r@{.}l    r@{.}l@{\,$\pm$\,}r@{.}l    r@{.}l@{\,$\pm$\,}r@{.}l     r@{.}l@{\,$\pm$\,}r@{.}l     r@{.}l@{\,$\pm$\,}r@{.}l    r@{.}l@{\,$\pm$\,}r@{.}l}
    \hline\hline
    $\dstdstpi$ & \multicolumn{4}{c}{$\zc(4020)^{-}\pip$} & \multicolumn{4}{c}{$D_1(2420)^{0}\dstzero$} & \multicolumn{4}{c}{$D_1(2420)^{+}D^{*-}$} & \multicolumn{4}{c}{$\mathrm{NR}\,D^{*0}$} \\\hline
    \hspace{0.2cm}$\zc(4020)^{-}\pip$         &  61&4&4&6  &  \multicolumn{2}{c}{ }-  & \multicolumn{4}{c}{ }- & \multicolumn{4}{c}{ }-  \\
    \hspace{0.2cm}$D_1(2420)^{0}\dstzero$     &  $-$6&4&3&5  &  10&9&3&6  & \multicolumn{2}{c}{ }- & \multicolumn{4}{c}{ }-   \\
    \hspace{0.2cm}$D_1(2420)^{+}D^{*-}$       & $-$1&2&2&3  & $-$5&0&3&3  &  10&0&3&3  & \multicolumn{2}{c}{ }-   \\
    \hspace{0.2cm}$\mathrm{NR}\,D^{*0}$       & $-$4&2&2&0  &  0&0&0&0  & $-$4&7&2&5  &  39&3&4&2   \\
    \hline\hline
    $\pipi\jpsi$  &  \multicolumn{4}{c}{$\zc(4020)^{-}\pi^{+}$} & \multicolumn{4}{c}{$\zc(3900)^{-}\pi^{+}$} & \multicolumn{4}{c}{$\zc(4020)^{+}\pi^{-}$} & \multicolumn{4}{c}{$\zc(3900)^{+}\pi^{-}$} & \multicolumn{4}{c}{$(\pi^{+}\pi^{-})_\mathrm{S-wave}J/\psi$} & \multicolumn{4}{c}{$f_{2}J/\psi$} \\\hline
    \hspace{0.2cm} $\zc(4020)^{-}\pi^{+}$  &  5&1&1&5  & \multicolumn{2}{c}{ }- & \multicolumn{4}{c}{ }- & \multicolumn{4}{c}{ }- & \multicolumn{4}{c}{ }- &  \multicolumn{4}{c}{ }\hspace{6.9em}-\\
    \hspace{0.2cm} $\zc(3900)^{-}\pi^{+}$  &  0&7&1&1  &  11&3&2&4  & \multicolumn{2}{c}{ }- & \multicolumn{4}{c}{ }- & \multicolumn{4}{c}{ }- &  \multicolumn{4}{c}{ }\hspace{6.9em}-\\
    \hspace{0.2cm} $\zc(4020)^{+}\pi^{-}$  &  0&2&0&1  &  $-$0&3&0&2  &  5&1&1&5  & \multicolumn{2}{c}{ }-  & \multicolumn{4}{c}{ }- &  \multicolumn{4}{c}{ }\hspace{6.9em}-\\
    \hspace{0.2cm} $\zc(3900)^{+}\pi^{-}$  & $-$0&3&0&2  &   0&6&0&3  &  0&7&1&1  &  11&3&2&4  & \multicolumn{2}{c}{ }- &  \multicolumn{4}{c}{ }\hspace{6.9em}-\\
    \hspace{0.2cm} $(\pi^{+}\pi^{-})_\mathrm{S-wave}J/\psi$  &  4&2&1&5  & $-$0&8&2&2  &  4&2&1&5 & $-$0&8&2&2  &  49&0&15&0  & \multicolumn{4}{c}{\hspace{0.55em}-}\\
    \hspace{0.2cm} $f_{2}J/\psi$  &  0&3&0&5  & $-$1&6&1&0  &  0&3&0&5  & $-$1&6&1&0  &  0&0&0&0  & \multicolumn{4}{c}{$12.5\pm3.1$} \\
    \hline\hline
    $\pipi\hc$  &  \multicolumn{4}{c}{$\zc(4020)^{-}\pi^{+}$} & \multicolumn{4}{c}{$\zc(4020)^{+}\pi^{-}$} & \multicolumn{4}{c}{$(\pi^{+}\pi^{-})_\mathrm{S-wave}h_{c}$} \\ \hline
    \hspace{0.2cm}     $\zc(4020)^{-}\pi^{+}$  & 33&5&4&7  & \multicolumn{2}{c}{ }- &  \multicolumn{4}{c}{ }- \\
    \hspace{0.2cm}     $\zc(4020)^{+}\pi^{-}$  &  0&3&0&1  &  33&5&4&7  &  \multicolumn{2}{c}{ }- \\
    \hspace{0.2cm}     $(\pi^{+}\pi^{-})_\mathrm{S-wave}h_{c}$  &  10&5&2&4  &  10&5&2&4  &  11&7&8&1   \\ \hline \hline
    \end{tabular}
\end{center}
\end{table*}

\begin{table*}[htbp]
\caption{Summary of the production cross sections (in pb) of different $\zc(4020)^-$ decay channels in $\ee\to\zc(4020)^{-}\pip+c.c$. (The values for $\pim\jpsi$ channel are scaled by a factor of 100.)}
\label{tab:summary_xs}
\begin{center}
\begin{tabular}{l c@{\,$\pm$\,}c@{\,$\pm$\,}l c@{\,$\pm$\,}c@{\,$\pm$\,}l}
\hline\hline
\multicolumn{1}{l}{Production cross section}  & \multicolumn{3}{c}{4.395 GeV} & \multicolumn{3}{c}{4.416 GeV}\\ \hline
$\sigma^B(\ee\to\zc(4020)^-\pip+c.c.)\cdot \mathcal{B}[\zc(4020)^{-}\to\dstzero\dstminus]$\quad\quad   & 335 & 21 & 35 & 336 & 26 & 28\\
$\sigma^B(\ee\to\zc(4020)^-\pip+c.c.)\cdot \mathcal{B}[\zc(4020)^{-}\to\pim\jpsi] (\times100)$         & 108  & 25 & 48 & 139 & 30 & 55\\
$\sigma^B(\ee\to \zc(4020)^-\pip+c.c.)\cdot \mathcal{B}[\zc(4020)^{-}\to\pim\hc]$                      & 30  & 4  & 8  & 30  & 3  & 6 \\
\hline\hline
\end{tabular}
\end{center}
\end{table*}

%%%%%%%%%%%%%%%%%%%%%%%%%%%%%%%%%%%%%%%%%%%%%%%%%%%%%%%%%%%%%%%%
%%%%%%%%%%%%%%%%%     JP    Part             %%%%%%%%%%%%%%%%%%%
%%%%%%%%%%%%%%%%%%%%%%%%%%%%%%%%%%%%%%%%%%%%%%%%%%%%%%%%%%%%%%%%
\subsubsection{B. Spin and parity}
\label{sec:jptest}
In addition to the $1^+$ assignment for the $\zc(4020)^{-}$ state, which is favored by the nominal fit, other spin-parity hypotheses ($1^{-}$, $2^{+}$, and $2^{-}$) require further testing. Note that the $0^+$ assignment is forbidden by spin-parity conservation in $\ee\to\gamma^*\to\zc(4020)^{-}\pip$, and $0^-$ is prohibited by $\zc(4020)^{-}\to\pim\hc$. The amplitudes are reconstructed by replacing the $\zc(4020)^{-}(1^+)$ component with the $1^-$, $2^{+}$, or $2^{-}$ assumption in the PWA fit.
%The negative log-likelihood (NLL) values of these alternative fits are summarized in the methodology section {\bf E}.
The fit with the $1^+$ hypothesis yields the minimum NLL value, which indicates the best spin-parity assignment for the resonance $\zc(4020)^{-}$. The fit quality deteriorates when considering alternative $\jp$ hypotheses.

Furthermore, to test the spin-parity of the $\zc(4020)^{-}$ as $1^+$ against other $J^P$ assignments ($1^-$, $2^+$, and $2^-$), two hypotheses are examined~\cite{Zhu:2006pfm}: the null hypothesis $H_0$ and the alternative hypothesis $H_1$.
Under $H_0$, data are assumed to be consistent with a single $\zc(4020)^{-}$ component with one of $\jp$ assignment: $1^-$, $2^+$, or $2^-$, while the $H_1$ posits that data can be better described with an additional $1^+$ $\zc(4020)^{-}$ component besides the $H_0$ components.
Following the fit, the significance of the $1^+$ assignment relative to other hypotheses is assessed by examining $\Delta(-2\ln\mathcal{L})$ as a function of $\Delta(\mathrm{ndof})$ to estimate the $p$-value and its corresponding significance level. 
The detailed differences are summarized in the Table~\ref{tab:S2_jp} in the methodology section \textbf{E}.
Comparing the hypotheses $H_0$ and $H_1$, we obtain confidence levels of $18.0\sigma$, $18.7\sigma$, and $15.8\sigma$ to exclude the hypotheses with only $\jp=1^{-}$, $2^{+}$, and $2^{-}$, respectively.

\begin{figure*}[htbp]
\begin{center}
\includegraphics[width=0.3\linewidth]{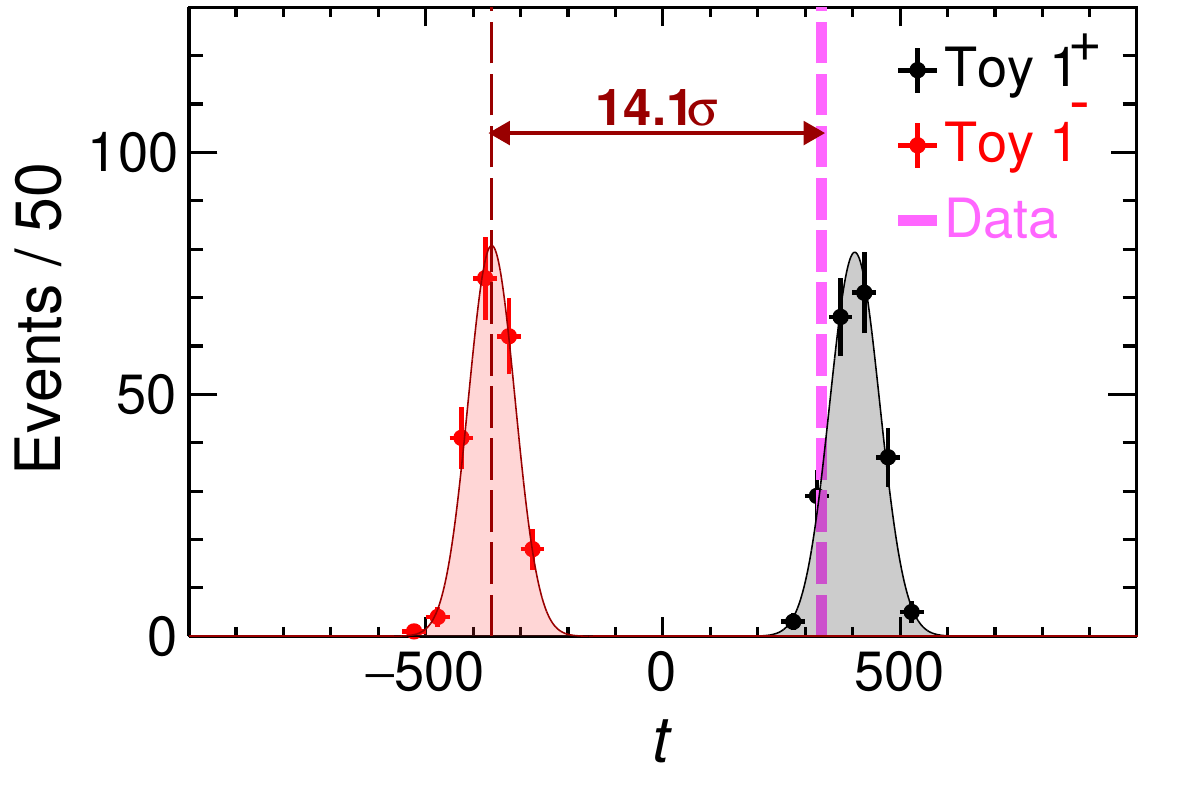}
\includegraphics[width=0.3\linewidth]{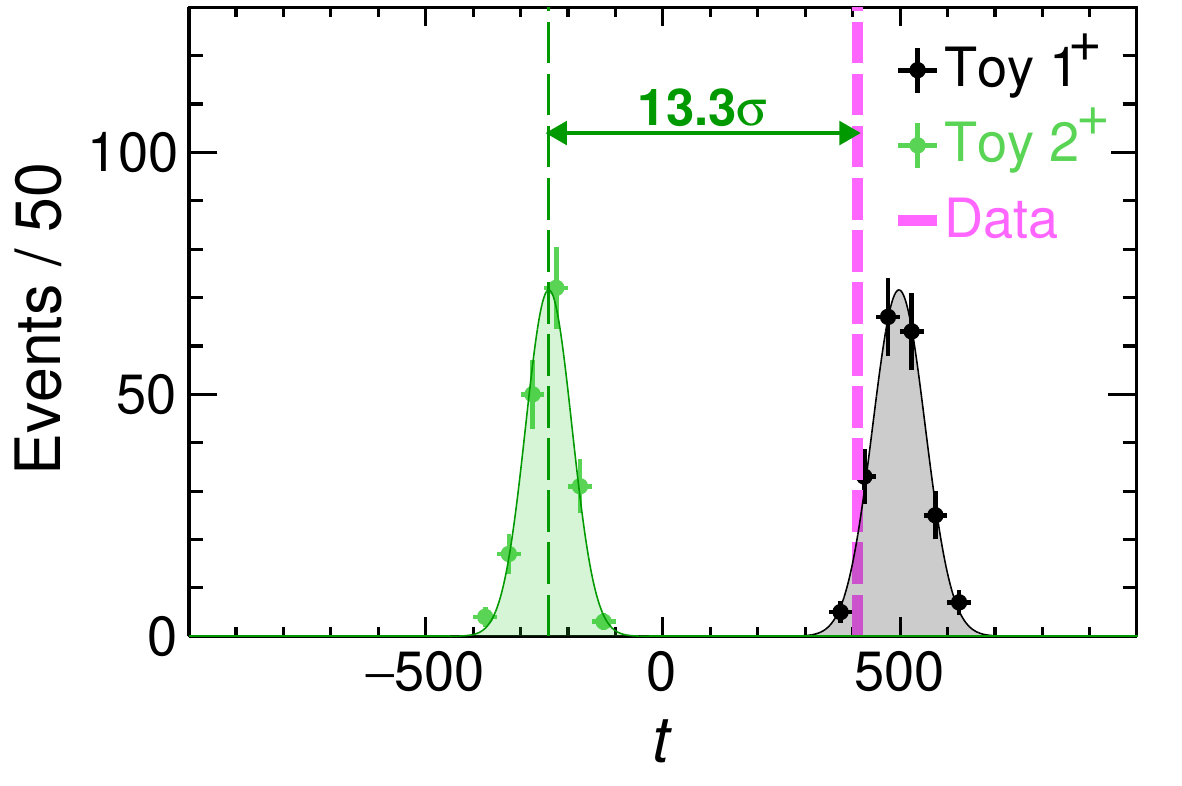}
\includegraphics[width=0.3\linewidth]{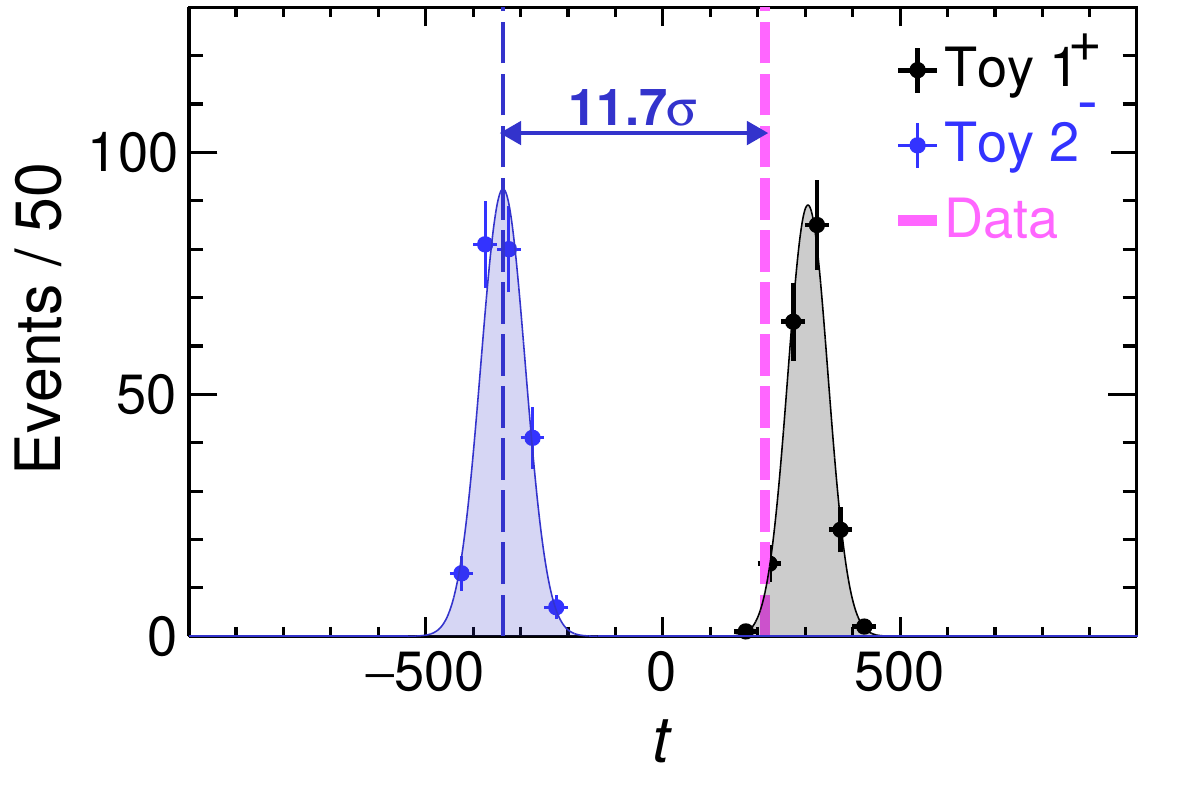}
\caption{The distributions of the likelihood ratio $t\equiv-2[-\ln L^{1^+}-(-\ln L^{J^P})]$. The black dots denote the toy MC samples generated under the hypothesis of $J^P=1^+$. The red, green, and blue dots represent the samples generated under the hypotheses $J^P=1^-,2^+$, and $2^-$, respectively. The pink-dashed vertical lines show the $t$ values in the data, and the deviations between the toy MC samples and the data are noted on the plots.}
\label{fig:t-jp-finalsys}
\end{center}
\end{figure*}

To consider the statistical fluctuation effect in estimating the $\jp$ significance, an additional test is conducted using a pseudo-experiment based on MC simulations, in which variations according to systematic uncertainties are taken into account.
A large set of toy MC samples is generated according to the fitted amplitude model, then subjected to detector response and the same event selection criteria as applied to real data.
In each toy MC sample, the yields of signals and backgrounds are set to those observed in data, with fluctuations according to the Poisson distributions.  Based on the NLL values of the PWA fits under different $\zc(4020)^{-}$ $\jp$ assumptions, the likelihood ratio variable $t\equiv\Delta(-2\ln\mathcal{L})= -2\ln\mathcal{L}^{\jp}-(-2\ln\mathcal{L}^{1^+})$ is used to evaluate the discrimination level.
The $t$ distributions for various $\jp$ hypotheses are illustrated in Fig.~\ref{fig:t-jp-finalsys}. In each plot, the toy MC samples in the right peak are generated with the amplitude model under the $\jp=1^+$ hypothesis, while the remaining samples are generated assuming $\jp=1^{-}$, $2^{+}$, or $2^{-}$. It is evident that all $t$ values from nominal fit to data favor the toy samples generated under the $\jp=1^+$ hypothesis.
The $t$ distributions can be well-fitted with Gaussian functions with parameters $(\overline{t},\,\sigma_{t})$.
The deviations between data and the toy samples assuming $\jp=1^{-}$, $2^{+}$, and $2^{-}$ are calculated as $(t_{\mathrm{data}}-\overline{t})/\sigma_{t}$, yielding confidence levels of rejection power of $14.1\sigma$, $13.3\sigma$, and $11.7\sigma$, respectively. Therefore, the spin-parity of $\zc(4020)^-$ is established to be $1^+$.

%%%%%%%%%%%%%%%%%%%%%%%%%%%%%%%%%%%%%%%%%%%%%%%%%%%%%%%%%%%%%%%%
%%%%%%%%%       systematic uncertainty  Part        %%%%%%%%%%%%
%%%%%%%%%%%%%%%%%%%%%%%%%%%%%%%%%%%%%%%%%%%%%%%%%%%%%%%%%%%%%%%%
\subsection{V. SYSTEMATIC UNCERTAINTIES}

The sources of systematic uncertainties in this analysis can be divided into two categories: the three-body cross sections, which affect only the FFs, and those from the PWA, which affect both the FFs and the pole positions.

For the systematic uncertainty from three-body cross sections, those associated with cross section measurement include tracking, particle identification (PID), photon detection, ISR factor, integrated luminosity, cited BFs, MC model, and fitting. These relevant systematic uncertainties of three processes are quoted directly from Refs.~\cite{BESIII:2023cmv,BESIII:2016bnd,BESIII:2016adj}, as 7.3\%, 5.8\%, 16.3\% at $\sqrt{s}=4.395\gev$ and 7.0\%, 5.8\%, 16.3\% at $\sqrt{s}=4.416\gev$.

The systematic uncertainties associated with the pole position of the $\zc(4020)^{-}$ and the FFs arise predominantly from the model dependencies of the $\pi\pi$ S-wave amplitude, the background estimation, the fit model, $\zcone(3900)^{\pm}$ propagator, other fixed parameters, the fit strategy and the mass resolution. For each
viable term, the difference between the nominal and the alternative fit results is taken as the systematic uncertainty.  
The uncertainty of $\pi\pi$ S-wave amplitude is estimated by replacing these amplitudes with $\pi\pi$ scattering amplitudes, which have been used in the soft $\pipi$ mass spectrum analyses by CLEO~\cite{CLEO:2011upl} and BESIII~\cite{BESIII:2016tqo}. The uncertainty from the background estimation is determined by altering the sideband regions and refitting. 
The modified $\jpsi$ mass sideband regions in $\pipijpsi$ are $(3.020,\,3.050)$ or $(3.150,\,3.180)\gevcc$, and the modified $\hc$ mass sideband regions in $\pipihc$ are $(3.470,\,3.490)$ or $(3.560,\,3.580)\gevcc$. For the $\dstdstpi$ channel, the background-free analysis is performed under nominal conditions, and the few background events survived in the hadron MC sample are adopted to check the systematic uncertainty. The uncertainty from the fit model is studied by adding the NR component with highest significance out of the baseline ($\mathrm{NR}(\dstminus\pip)^{2^{+}}\dstzero$, $\mathrm{NR}(\pim\jpsi)^{2^{-}}\pip$, $\mathrm{NR}(\pim\hc)^{2^{+}}\pip$) to check their influence. 
The contribution of $D_1(2420)$ at $\sqrt{s}=4.395\gev$ is also investigated. The uncertainty from fit model variations is quantified by combining two independent sources in quadrature. $\zc(3900)$ is described by a Flatte-type BW in baseline, which is varied to a constant width BW~\cite{BESIII:2025qkn} as the alternative. The uncertainty of the introduced propagator parameters is estimated by changing the fixed parameters within one standard deviation individually and summing up the contributions in quadrature. 
Potential fitting biases in the PWA fit are extensively tested with toy MC studies, and the differences between the generation truths and the fitted outcomes are taken as the systematic uncertainties from the fit strategy. The detector resolution is not considered in the nominal fit, and the resulting uncertainty on the pole width of the $\zc(4020)^{-}$ is evaluated by comparing the input width in the simulation and the reconstructed one, considering the detector responses.

The systematic uncertainties of the extracted pole positions and FFs are summarized in Table~\ref{tab:sys_pole} and Table~\ref{tab:sys_FF}, respectively, where the total systematic uncertainty is the quadratic sum of the individual contributions.
The influence of the relevant systematic uncertainties on the significance of alternative spin-parity assignments is studied as well in the PWA fit, and the significances of $1^+$ over other $J^P$ assignments are all above $14.1\sigma$. Detailed numbers are summarized in Table~\ref{tab:sys_on_jp}.

\begin{table*}[htbp]
 \caption{The absolute systematic uncertainties of the pole mass ($m_\text{pole}$, in $\mevcc$) and width ($\Gamma_\text{pole}$, in $\mev$) for the $\zc(4020)^{-}$ state.}
	\label{tab:sys_pole}
	\begin{center}
	\begin{tabular}{lcccccccc}
	\hline\hline
	& \multicolumn{2}{c}{Pole\,1} & \multicolumn{2}{c}{Pole\,2} & \multicolumn{2}{c}{Pole\,3}& \multicolumn{2}{c}{Pole\,4}\\ \hline
	Source    &$m_\text{pole}$   & $\Gamma_\text{pole}$  &$m_\text{pole}$   & $\Gamma_\text{pole}$&$m_\text{pole}$   & $\Gamma_\text{pole}$&$m_\text{pole}$   & $\Gamma_\text{pole}$\\\hline
    $\pi\pi$ S-wave           & 0.38 & 0.77 & 0.29 & 0.51 & 0.08 & 1.95 & 0.00 & 1.83 \\
	Background                & 0.05 & 0.17 & 0.06 & 0.21 & 0.10 & 0.08 & 0.09 & 0.05 \\
	Fit model                 & 0.40 & 1.09 & 0.18 & 1.60 & 0.79 & 7.04 & 0.55 & 1.07 \\
	$\zc(3900)$ propagator    & 0.03 & 1.08 & 0.14 & 0.52 & 0.38 & 5.57 & 0.53 & 5.16 \\
	Other BW parameters       & 0.36 & 1.43 & 0.28 & 1.23 & 0.70 & 10.98& 0.70 & 9.15 \\
	Fit strategy              & 1.07 & 1.33 & 0.86 & 0.80 & 0.09 & 9.07 & 0.14 & 8.49 \\
	Mass resolution           & -    & 1.37 & -    & 1.37 & -    & 1.37 & -    & 1.37 \\
	\hline
	Total                     & 1.26 & 2.95  & 0.98 & 2.68 & 1.13 & 17.00 & 1.05 & 13.74 \\
	\hline \hline
	\end{tabular}
	\end{center}
\end{table*}

\begin{table*}[htbp]
\caption{The relative systematic uncertainties (\%) of FFs in different $\zc(4020)^{-}$ decay channels. }
\label{tab:sys_FF}
\begin{center}
\begin{tabular}{lccccccc}
\hline\hline
Data set & \multicolumn{3}{c}{$4.395\gev$} & & \multicolumn{3}{c}{$4.416\gev$} \\
Decay mode  &$\dstzero\dstminus$ & $\pim\jpsi$ &$\pim\hc$ & &$\dstzero\dstminus$ & $\pim\jpsi$ &$\pim\hc$ \\
\hline
     $\pi\pi$ S-wave           & 1.0  & 19.0  & 0.4    & & 0.8 & 12.3  & 2.0  \\
     Background                & 0.7  & 8.0   & 6.6    & & 0.4 & 6.8   & 6.3  \\
     Fit model                 & 7.0  & 20.0  & 12.3   & & 3.5 & 16.5  & 3.5  \\
     $\zcone(3900)$ propagator & 0.1  & 13.4  & 2.5    & & 0.9 & 12.0  & 3.7  \\
     Other BW parameters       & 1.2  & 29.9  & 12.3   & & 1.9 & 28.0  & 11.8 \\
     Fit strategy              & 2.5  & 8.2   & 2.5    & & 1.3 & 10.7  & 0.2  \\
     \hline
     Total                     & 7.6  & 44.3  & 19.0   & & 4.4 & 38.9  & 14.4  \\
	\hline\hline
	\end{tabular}
    \end{center}
\end{table*}

\begin{table*}[!htbp]
\caption{Confidence levels of $H_1$ hypothesis versus $H_0$ hypothesis for the $\zc(4020)^-$ spin-parity assignments.}
\label{tab:sys_on_jp}
\begin{center}
\begin{tabular}{lcccccc}
\hline \hline
Source & $1^{+}$over $1^{-}$ & $1^{+}$over $2^{+}$ & $1^{+}$over $2^{-}$ & $1^{-}$over $1^{+}$ & $2^{+}$over $1^{+}$ & $2^{-}$over $1^{+}$\\ \hline
  Default                               &$18.0\sigma$   &$18.7\sigma$   &$15.8\sigma$   &$1.5\sigma$   &$1.5\sigma$   &$3.6\sigma$\\
  $\pipi$ S-wave                        &$18.7\sigma$   &$19.2\sigma$   &$18.3\sigma$   &$0.5\sigma$   &$0.0\sigma$   &$3.9\sigma$\\
  Background                            &$17.2\sigma$   &$17.8\sigma$   &$15.7\sigma$   &$1.6\sigma$   &$1.4\sigma$   &$3.8\sigma$\\
  Fit model                             &$17.2\sigma$   &$18.5\sigma$   &$14.1\sigma$   &$3.1\sigma$   &$0.5\sigma$   &$3.9\sigma$\\
  $\zcone(3900)$ BW                     &$17.8\sigma$   &$18.8\sigma$   &$15.4\sigma$   &$1.9\sigma$   &$2.0\sigma$   &$3.2\sigma$\\
  Others BW parameters                  &$17.5\sigma$   &$18.7\sigma$   &$15.4\sigma$   &$1.1\sigma$   &$1.6\sigma$   &$3.7\sigma$\\
\hline \hline
\end{tabular}
\end{center}
\end{table*}

\subsection{VI. SUMMARY}
A multi-channel joint analysis of the $\zc(4020)^{-}$ state is performed for the first time using data samples collected with the BESIII detector at the BEPCII collider. The data were taken at center-of-mass energies of $\sqrt{s}=4.395$ and $4.416\gev$, with a total integrated luminosity of $1598.9\invpb$. The analysis is based on the processes $\ee\to\dstdstpi$, $\pipijpsi$, and $\pipihc$.
The spin-parity of the $\zc(4020)^{-}$ is determined for the first time to be $1^+$ with the significance larger than $11.7\sigma$. The pole mass and width of the $\zc(4020)^{-}$ are determined to be $m_\mathrm{pole1}=(4022.44\pm1.55\pm1.26)\mevcc$, $\Gamma_\mathrm{pole1}=(38.54\pm2.94\pm2.95)\mev$ and $m_\mathrm{pole2}=(4023.01\pm1.35\pm0.98)\mevcc$, $\Gamma_\mathrm{pole2}=(35.02\pm2.20\pm2.68)\mev$.
Based on the fit, the $\zc(4020)^-$ Born cross sections times the corresponding BFs are obtained and summarized in Table~\ref{tab:summary_xs}.
The ratio of relative BFs of $\zc(4020)^{-}$ decaying into $\dstzero\dstminus$, $\pim\jpsi$, and $\pim\hc$ are derived for the first time as $\mathcal{B}[\zc(4020)^{-}\to\pim\jpsi]/\mathcal{B}[\zc(4020)^{-}\to\dstzero\dstminus]=(3.6\pm0.6\pm1.6)\times10^{-3}$ and $\mathcal{B}[\zc(4020)^{-}\to\pim\hc]/\mathcal{B}[\zc(4020)^{-}\to\dstzero\dstminus]=(8.9\pm1.3\pm2.3)\times10^{-2}$, respectively. For these results, the first uncertainties are statistical, and the second are systematic. In these results, the $\zc(4020)^{-}$ exhibits a significantly stronger coupling to $D^{*}\bar{D}^{*}$ process than to hidden-charm channels.

With the determination of $J^P=1^+$, the notation of the $\zc(4020)^{-}$ can be updated to $\zcone(4020)^-$ following the PDG naming scheme.
The alignment of its spin-parity with that of the $\zcone(3900)$, combined with the relative BF results, strongly suggests both $\zcone$ being $D^{*}\bar{D}^{(*)}$ molecules' structure rather than a charmonium core.

Eight poles are investigated and searched on the eight-sheet Riemann surface due to three branch points, as summarized in Table~\ref{tab:S1_pole}.
Several of them (Poles 2, 5-8) reside on non-adjacent Riemann sheets relative to the physical region~\cite{Yamada:2022xam}, whose contributions are suppressed but non-negligible.
Additional experimental data are necessary to further constrain the above estimation, and more theoretical research is required to quantify the relative contributions of the distinct poles.

\subsection{ACKNOWLEDGMENTS}
\input{acknowledgement_2025-03-21.tex}

\subsection{VII. METHODOLOGY}

\subsubsection{A. Event selections}
\label{sec::methods:A}
The charged tracks and photon candidates are identified based on the criteria outlined in Ref.~\cite{BESIII:2021dmo}. Candidate events for the three signal processes are selected as follows.

\subsubsection{1. $\ee\to\dstdstpi$}
Two decay channels of $\dstzero$ in the signal processes $\ee\to\dstdstpi$ are considered: $\dstzero\to\dzero\pizero$ or $\dzero\gamma$, while $\dstminus\to\dzerobar\pim$.
We utilize a partial reconstruction method to reconstruct those decays, in which $\pim$ originated from $\dstminus$ decays are not reconstructed because of their low momentum and low detection efficiency.

Both $\dzero$ and $\dzerobar$ are reconstructed via three dominant channels: $\dzero\to\kaonm\pip$, $\kaonm\pip\pizero$, and $\kaonm\pip\pip\pim$.
The reconstructed invariant masses are required to fall within the range of $(1.850,\,1.880)\gevcc$. The charged kaons and pions from the  $\dzero$ ($\dzerobar$) decays are identified by comparing PID likelihoods, which are formed by combining measurements of the energy deposited in the MDC~(d$E$/d$x$) and the flight time in the TOF based on the kaon and pion hypotheses, $\mathcal{L}(K)>\mathcal{L}(\pi)$ and $\mathcal{L}(\pi)>\mathcal{L}(K)$, respectively.
Then, one $\pizero$ or $\gamma$ is detected and associated with the reconstructed $\dzero$.
The $\pizero$ from the $\dstzero\to\pizero\dzero$ decay is reconstructed by two photons with an invariant mass within the interval $(0.120,\,0.145)\gevcc$.
In addition, the charged $\pip$ originating from the primary production vertex, labelled as bachelor $\pip$, are identified with the PID requirement by looping over all charged tracks.
The reconstructed $\dzero\pip$ invariant mass is required to be $MQ(\dzero\pip)>2.03\gevcc$ to reject the $\pip$ from $\dstplus$ decay.
Here and in the following, a resolution decorrelation procedure is applied to improve the resolution of the reconstructed invariant masses:
for the invariant mass reconstructed by particles $A$ and $B$, one can decorrelate the instrumental resolution effect of the intermediate state $B$ by adopting the variables $MQ(AB)=M(AB)-M(B)+m(B)$ and $RQ(AB)=RM(AB)+M(B)-m(B)$, where $M$, $RM$, and $m$ are the invariant mass, the recoil invariant mass, and known mass, respectively.
$B$ represents the intermediate states ($\dzero$, $\dzerobar$, and $\pizero$).

\begin{figure*}[htbp]
\begin{center}
    \includegraphics[width=0.32\linewidth]{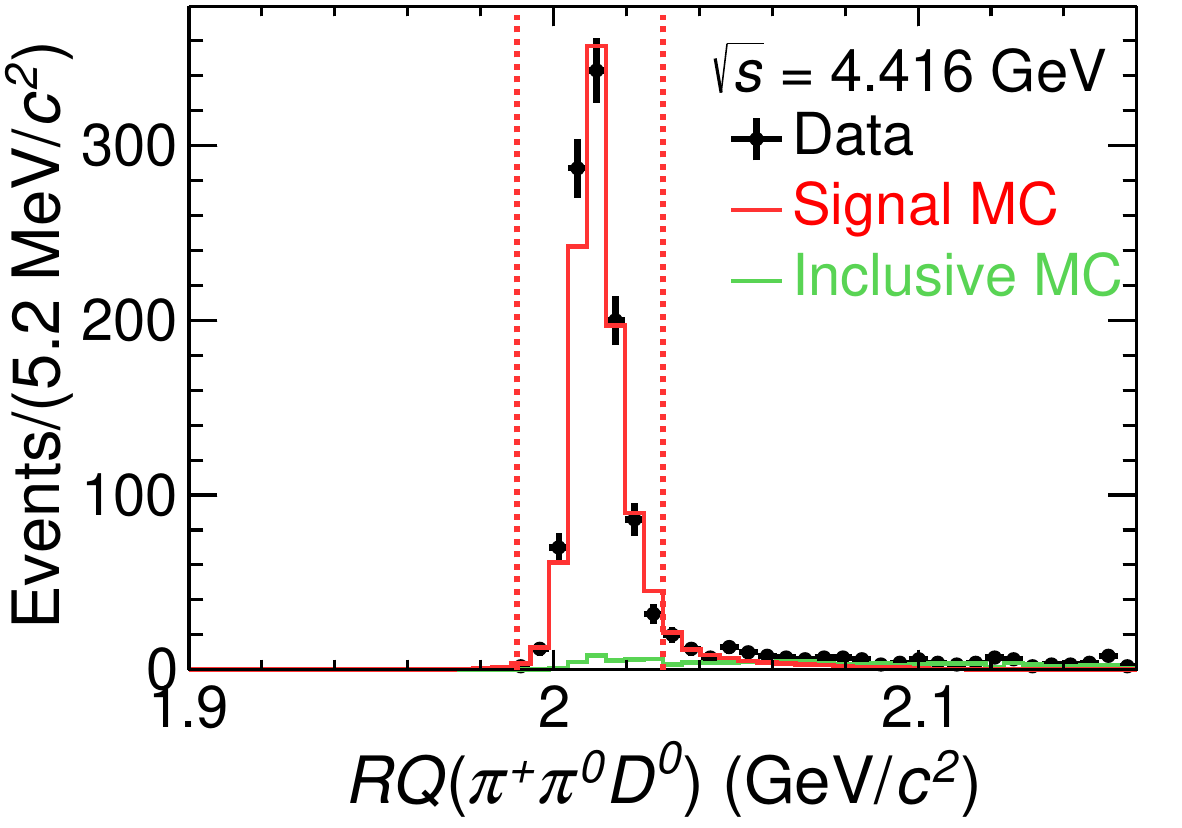}
    \includegraphics[width=0.32\linewidth]{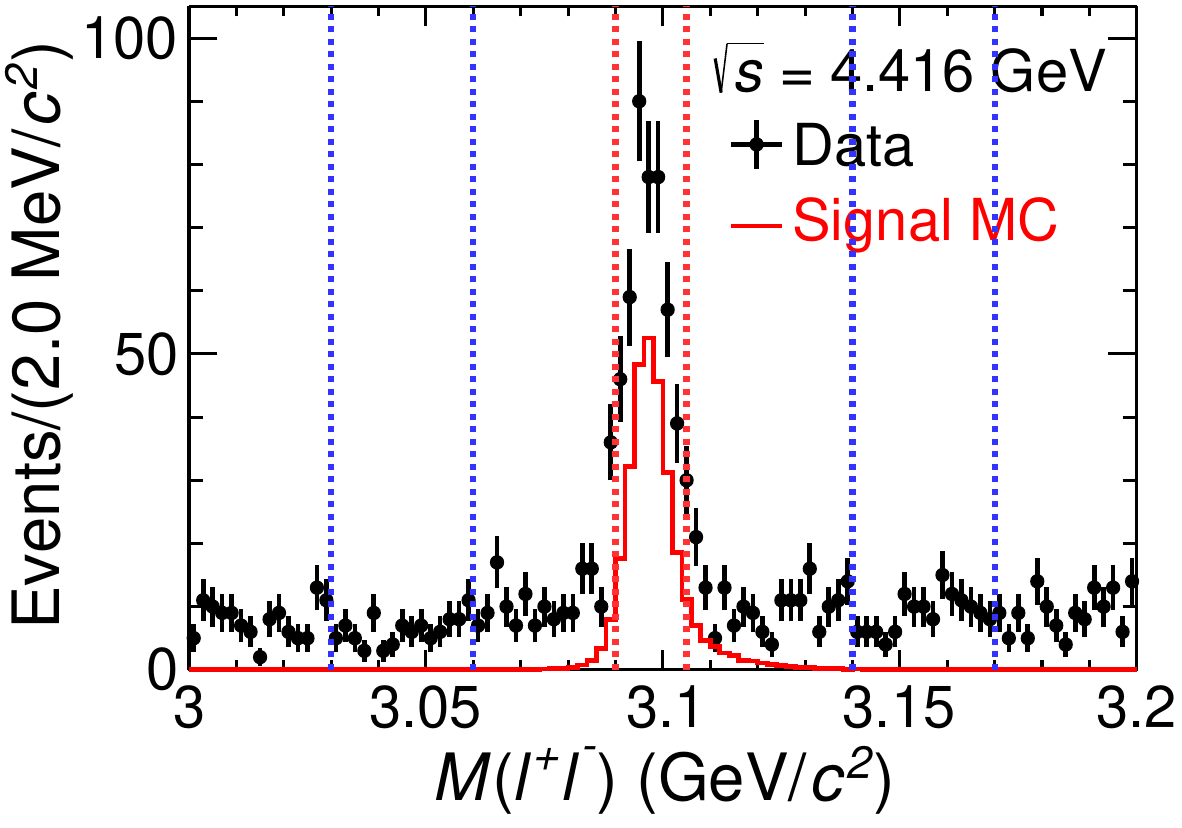}
    \includegraphics[width=0.32\linewidth]{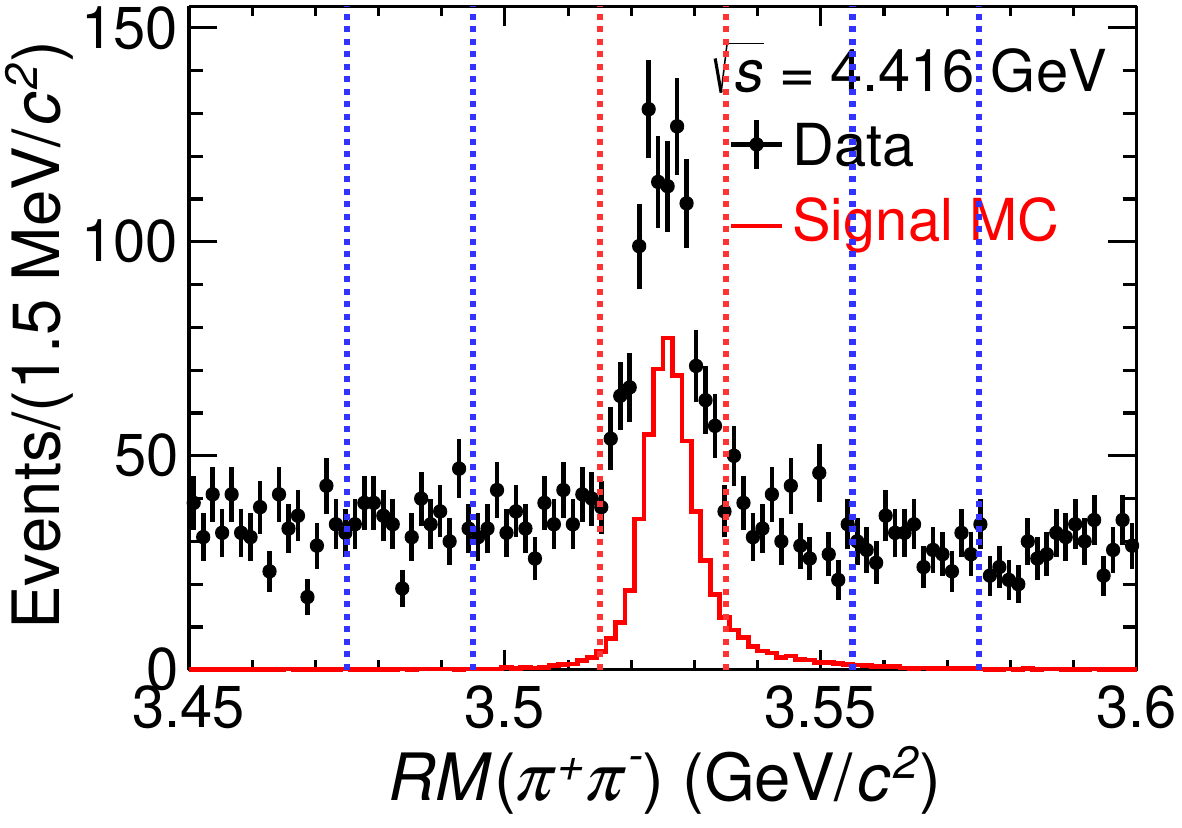}
    \caption{At $\sqrt{s}=4.416\gev$, the left plot shows the $RQ(\dzero\pizero\pip)$ distribution in the process $\ee\to\dstdstpi$, where backgrounds are negligible as indicated with the tiny amount of survived inclusive MC events. The middle plot shows the $M(\lp\lm)$ distribution in $\ee\to\pipijpsi$ and the right plot presents the $RM(\pip\pim)$ distribution in $\pipihc$. The signal and sideband regions are marked with red and blue vertical dotted lines, respectively.}
    \label{fig:data_vs_mc}
\end{center}
\end{figure*}

The $\dstzero$ candidates are selected by requiring an invariant mass window of $MQ(\pizero\dzero)\in(2.004,\,2.009)\gevcc$ or $MQ(\gamma\dzero)\in(1.995,\,2.015)\gevcc$.
The signal candidates are identified by requiring the recoil invariant mass $RQ(\dstzero\dzerobar\pip)$ falling within $(0.120,\,0.160)\gevcc$.
Events with the recoil invariant mass $RQ(\dzero\pizero\pip)$ in the interval $(1.990,\,2.030)\gevcc$ or $RQ(\dzero\gamma\pip)$ in $(1.990,\,2.040)\gevcc$ are kept for further analysis.

For the events that meet the aforementioned criteria, a kinematic fit is performed to constrain the reconstructed $\dzero$, $\dstzero$, and $\dzerobar$ to their respective known masses.
Additionally, the recoil mass of $\dstzero\dzerobar\pip$ and  $\dstzero\pip$ are constrained to the known mass of $\pim$ and $\dstminus$~\cite{ParticleDataGroup:2024cfk}, respectively.
For the $\dstzero\to\dzero\pizero$ decay, an additional mass constraint is applied to the reconstructed invariant mass of $\pizero$.
If multiple candidates survive in a single event, the one with the minimum $\chi^{2}$ value from the kinematic fit is retained.

After applying all the selection criteria outlined above, the background contribution is expected to be negligible, as illustrated by the $RQ(\dzero\pizero\pip)$ distribution depicted in the left panel of Fig.~\ref{fig:data_vs_mc}. Therefore, we perform a nearly background-free analysis in the case of the $\ee\to\dstdstpi$ process, and the influence of the backgrounds is treated as a systematic uncertainty.

\subsubsection{2. $\ee\to\pipijpsi$}
With the same event selection criteria in Ref.~\cite{BESIII:2016bnd}, the $\jpsi$ candidates are reconstructed by $\lp\lm(l=e,\mu)$ pairs.
The number of selected charged tracks in each event must be equal to four for the signal process $\pip\pim\lp\lm$.
Charged tracks with momentum in the MDC less than $1.06\gevc$ are assumed to be pions, while the others are considered as leptons.
The leptons with deposited energy in the EMC, $E_\mathrm{EMC}<0.35\gev$ are classified as muons, while those with $E_\mathrm{EMC}>1.1\gev$ are treated as electrons.

To remove the backgrounds from radiative Bhabha and radiative dimuon events with $\ee$ from $\gamma$-conversion misidentified as $\pipi$ candidates, the cosine of the angle between $\pip$ and $\pim$ in the laboratory system $\cos\theta_{\pi^{+}\pi^{-}}$is required to be less than $0.98$.
The four-momentum (4C) kinematic fit is performed to constrain the momentum of $\pip\pim\lp\lm$ to the initial $\ee$ system with a fit quality requirement $\chi^{2}_{\rm 4C}<40$ to improve the momentum resolution and reduce the background.
Moreover, for the $\jpsi\to\ee$ process, the constraint of the cosine of the angle between pion and electron $\cos\theta_{\pi^{\pm}e^{\mp}}<0.98$ is applied to reject background from $\gamma$-conversion.
To improve the purity of $\jpsi\to\mu^{+}\mu^{-}$ decays, we require that the number of hit layers in the muon chamber must be larger than 5 for at least one muon candidate.

The events that remain within the mass window of $M(\lp\lm)\in(3.090,\,3.105)\gevcc$ are classified as signal candidates. The background events in the signal region are estimated using the sideband range of the $\jpsi$ invariant mass spectrum, defined as $M(\lp\lm)\in(3.030,\,3.060)\cup(3.140,\,3.170)\gevcc$, as shown in the middle panel of Fig.~\ref{fig:data_vs_mc}.
Finally, the four-momentum updated from kinematic 4C with additional mass constraint of $\jpsi$ is adopted to further the PWA fit.

\subsubsection{3. $\ee\to\pipihc$}
The event selection criteria for this channel are consistent with Ref.~\cite{BESIII:2016adj}.
For the $\ee\to\pipihc$ process, where $\hc$ decays into $\gamma\etac$, the $\etac$ is reconstructed through 16 different hadronic decay processes, $\ie$, $\ppbar$, $2(\pipi)$, $2(\kk)$, $\pipi\kk$, $\pipi\ppbar$, $3(\pipi)$, $2(\pipi)\kk$, $\ks K^{\pm}\pi^{\mp}$, $\ks K^{\pm}\pi^{\mp}\pipi$, $\kk\pizero$, $\ppbar\pizero$, $\kk\eta$, $\pipi\eta$, $2(\pipi)\eta$, $\pipi\pizero\pizero$, and $2(\pipi)\pizero$.

For each event, the total net charge of all tracks is required to be zero, and the number of charged tracks is to be either 4, 6, or 8, depending on the specific decay channels of $\hc$.
For the decay channels with an intermediate state $\ks$, two charged tracks are constrained to originate from a common vertex and are required to have an invariant mass such that $|M(\pip\pim)-m_{\ks}|<20\mevcc$, where $m_{\ks}$ is the $\ks$ nominal mass~\cite{ParticleDataGroup:2024cfk}.
The decay length of the $\ks$ candidate must exceed twice its standard deviation.
The $\ks$ candidate with the smallest $\chisq_{\ks}$ of the mass-constraint fit is selected.
The $\pizero$ and $\eta$ meson candidates are reconstructed by photon pairs with invariant masses $M(\gamma\gamma)$ within $(0.110,\,0.150)\gevcc$ and $(0.500,\,0.570)\gevcc$, respectively.

The same 4C method used in the $\ee\to\pipijpsi$ process is applied to all final particles.
In cases where one event can be selected by different channels with the same number of charged tracks, the best candidate is chosen based on a combination of chi squares from the 4C kinematic fit, the PID of each charged track, and the mass-constrainted fit of $\pizero/\eta$, represented by $\chi^{2}=\chi^{2}_{\mathrm{4C}}+\chi^{2}_{\mathrm{PID}}+\chi^{2}_{\pizero/\eta}$. The candidate with the minimum overall $\chi^{2}$ value is retained for further analysis.

To suppress background, for final states consisting of only charged tracks, $RM(\pip\pim\gamma)$ is required to fall within the range $(2.934,\,3.034)\gevcc$, and the four-momentum kinematic fit must have $\chisq_{\rm 4C}<35$.
For final states including $\pizero/\eta$ candidates, we require $RM(\pip\pim\gamma)$ to be in the range $(2.939,\,3.029)\gevcc$, and the four-momentum kinematic fit must have $\chisq_{\rm 4C}<20$.
If there are multiple combinations of $\pip\pim$ from $\ee$ annihilations and $\gamma$ from $\hc$ decays in an event, the candidate with the minimum $|M(\eta_c)-m_{\etac}|$ is kept in each event, where $m_{\etac}$ is the $\etac$ nominal mass~\cite{ParticleDataGroup:2024cfk}.

The signal events are chosen with $RM(\pip\pim)$ falling within the $\hc$ signal mass window $(3.515,\,3.535)\gevcc$. The background events are estimated using the $\hc$ mass sidebands, with $RM(\pip\pim)$ in $(3.475,\,3.495)$ or $(3.555,\,3.575)\gevcc$, as shown in the right-hand side panel of Fig.~\ref{fig:data_vs_mc}.
Finally, the four-momentum updated from kinematic 4C with additional mass constraint of $\hc$ is adopted to further the PWA fit.

\subsubsection{B. Helicity amplitude}
\subsubsection{1. $\ee\to\dstdstpi$}
The decay into a three-body system is described as a sequential two-body decay involving the introduced intermediate state, with the corresponding assignments of its spin and parity.
The process $\ee\to\dstdstpi$ could occur through the $\zc$ and $\Ri$ states.
\begin{itemize}
    \item[1)] $\ee\to\gammas\to\zcm\pip, \zcm\to\dstzero\dstminus,\dstzero\to\dzero\pizero(\gamma), \dstminus\to\dzerobar\pim$,
    \item[2)] $\ee\to\gammas\to\Ri\dstzero, \Ri\to\dstminus\pip,\dstzero\to\dzero\pizero(\gamma), \dstminus\to\dzerobar\pim$,
    \item[3)] $\ee\to\gammas\to\Ri'\dstminus, \Ri'\to\dstzero\pip,\dstzero\to\dzero\pizero(\gamma), \dstminus\to\dzerobar\pim$.
\end{itemize}

\begin{figure}[htp]
\includegraphics[width=0.4\textwidth]{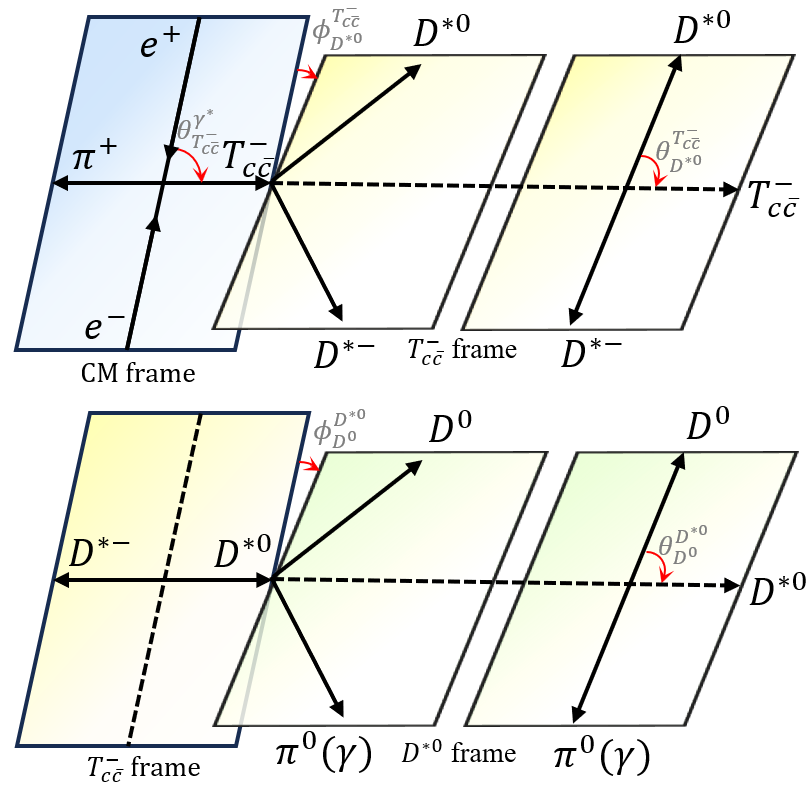}
\caption{Illustration of the subprocess $\ee\to\gammas\to\zcm\pip, \zcm\to\dstzero\dstminus,\dstzero\to\dzero\pizero(\gamma), \dstminus\to\dzerobar\pim$.}
\label{app:fig:decay-D-chain1}
\end{figure}

For the first subprocess $\ee\to\gammas\to\zcm\pip, \zcm\to\dstzero\dstminus, \dstzero\to\dzero\pizero(\gamma), \dstminus\to\dzerobar\pim$, we analyze the initial decay chain depicted in Fig.~\ref{app:fig:decay-D-chain1}, and its decay amplitude is described as:
\begin{widetext}
\begin{align}\label{app:eq:decay_amp_D1}
A_{D,1}(\lambda_{\gammas}, \lambda_{\pizero(\gamma)})=\sum_{\lambda_{\zcm},\lambda_{\dstzero},\lambda_{\dstminus}}A^{\gammas\to\zcm\pip}_{\lambda_{\zcm},0}R(m_{D^{*0}D^{*-}})A^{\zcm\to\dstzero\dstminus}_{\lambda_{\dstzero},\lambda_{\dstminus}}
 A^{\dstzero\to\dzero\pizero(\gamma)}_{0,\lambda_{\pizero(\gamma)}}A^{\dstminus\to\dzerobar\pim}_{0,0},
\end{align}
with
\begin{align}
&A^{\gammas\to\zcm\pip}_{\lambda_{\zcm},0}=H^{\gammas\to\zcm\pip}_{\lambda_{\zcm},0}D^{1*}_{\lambda_{\gammas},\lambda_{\zcp}}(\phi_{\zcm}^{\gamma^*},\theta_{\zcm}^{\gamma^*},0),\;\;A^{\zcm\to\dstzero\dstminus}_{\lambda_{\dstzero},\lambda_{\dstminus}}=H^{\zcm\to\dstzero\dstminus}_{\lambda_{\dstzero},\lambda_{\dstminus}}D^{J_{\zcm}*}_{\lambda_{\zcm},\lambda_{\dstzero}-\lambda_{\dstminus}}(\phi_{\dstzero}^{\zcm},\theta_{\dstzero}^{\zcm},0),\nonumber\\
&A^{\dstzero\to\dzero\pizero(\gamma)}_{0,\lambda_{\pizero(\gamma)}}=D^{1*}_{\lambda_{\dstzero},-\lambda_{\pizero(\gamma)}}(\phi^{\dstzero}_{\dzero},\theta^{\dstzero}_{\dzero},0),\;\;A^{\dstminus\to\dzerobar\pim}_{0,0}=D^{1*}_{\lambda_{\dstminus},0}(\phi^{\dstminus}_{\dzerobar},\theta^{\dstminus}_{\dzerobar},0),\nonumber
\end{align}
\end{widetext}
where the symbols $\lambda_{\gammas}$, $\lambda_{\zcm}$, $\lambda_{\dstzero}$, $\lambda_{\dstminus}$, and $\lambda_{\pizero(\gamma)}$ represent the helicities of the $\gammas$, $\zcm$, $\dstzero$, $\dstminus$, and $\pizero(\gamma)$ respectively.
The mother $\ee$ system is set as $\ee$ center-mass frame, the helicity angles $\phi_{\zcm}^{\gamma^*}$ and $\theta_{\zcm}^{\gamma^*}$ are defined as the azimuthal and polar angles of the $\zc^{-}$ momentum vector in the mother $\ee$ system.
Angle $\phi_{\dstzero}^{\zcm}$ is between the production and decay plane of $\zc^{-}$, while $\theta_{\dstzero}^{\zcm}$ is the angle between the $\dstzero$ momentum vector and the $\zcm$ momentum vector in the rest frame of the respective parent particle.
The $\phi_{\dzero}^{\dstzero}$ angle is the angle between the $\dstzero$ production plane and decay plane, and $\theta_{\dzero}^{\dstzero}$ is the angle between the $\dzero$ flight direction and the direction of the $\dstzero$. A similar definition applies to the decay of $\dstminus$. Figure~\ref{app:fig:decay-D-chain1} illustrates the involved helicity angles.
The symbol $R(m_{\dstzero\dstminus})$ represents the propagator for the resonance $\zcm$ (see Eq.~\ref{eq:zc4020}), and $J_{\zcm}$ denotes the spin of $\zcm$. The helicity amplitudes $H^{\dstzero\to\dzero\pizero(\gamma)}_{0,\lambda_{\pizero(\gamma)}}$ and $H^{\dstminus\to\dzerobar\pim}_{0,0}$ are set to 1 as normalization in the common decay in three subproceses, and the $\dst$ lineshape is modelled with a Breit-Wigner function and applied to the MC events accordingly.

\begin{figure}[h]
\begin{center}
\includegraphics[width=0.4\textwidth]{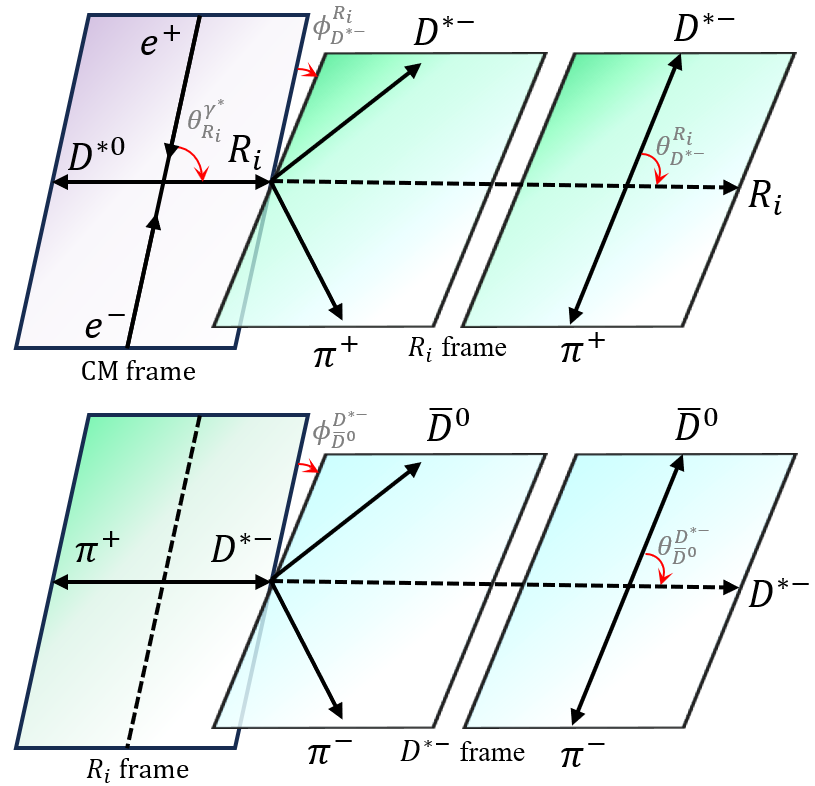}
\caption{The depicted diagram shows the subprocess of $\ee\to\gammas\to \Ri\dstzero, \Ri\to\dstminus\pip, \dstzero\to\dzero\pizero(\gamma), \dstminus\to\dzerobar\pim$.}
\label{app:fig:decay-D-chain2}
\end{center}
\end{figure}

For the second subprocess $\ee\to\gammas\to\Ri\dstzero, \Ri\to\dstminus\pip, \dstzero\to\dzero\pizero(\gamma), \dstminus\to\dzerobar\pim$, as illustrated in Fig.~\ref{app:fig:decay-D-chain2}, the amplitude is given by:
\begin{align}\label{app:eq:decay_amp_D2}
&A_{D,2}(\lambda_{\gammas},\lambda_{\pizero(\gamma)})\nonumber\\
=&\sum_{\Ri,\lambda_{\Ri},\lambda'_{\pizero(\gamma)}}A^{\gammas\to \Ri\dstzero}_{\lambda_{\Ri},\lambda_{\dstzero}}\,R(m_{D^{*-}\pi^+})
 A^{\Ri\to\dstminus\pip}_{\lambda_{\dstminus},0}\nonumber\\
&\times A^{\dstzero\to\dzero\pizero(\gamma)}_{0,\lambda'_{\pizero(\gamma)}}A^{\dstminus\to\dzerobar\pim}_{0,0}D^{J_{\pizero(\gamma)}*}_{\lambda'_{\pizero(\gamma)},\lambda_{\pizero(\gamma)}}(\tilde\phi_2,\tilde\theta_2,\tilde\gamma_2),
\end{align}
with
\begin{align}
A^{\gammas\to \Ri\dstzero}_{\lambda_{\Ri},\lambda_{\dstzero}}&=H^{\gammas\to \Ri\dstzero}_{\lambda_{\Ri},\lambda_{\dstzero}}D^{1*}_{\lambda_{\gammas},\lambda_{\Ri}-\lambda_{\dstzero}}(\phi_{\Ri}^{\gamma^*},\theta_{\Ri}^{\gamma^*},0),\nonumber\\
A^{\Ri\to\dstminus\pip}_{\lambda_{\dstminus},0}&=H^{\Ri\to\dstminus\pip}_{\lambda_{\dstminus},0}D^{J_{\Ri}*}_{\lambda_{\Ri},\lambda_{\dstminus}}(\phi_{\dstminus}^{\Ri},\theta_{\dstminus}^{\Ri},0),\nonumber
\end{align}
where the helicity angles $\phi_{\Ri}^{\gamma^*}$, $\theta_{\Ri}^{\gamma^*}$, $\phi_{\dstminus}^{\Ri}$ and $\theta_{\dstminus}^{\Ri}$ are defined similarly to the first subprocess, as depicted in Fig.~\ref{app:fig:decay-D-chain2}.
The term $R(m_{D^{*-}\pi^+})$ denotes the Breit-Wigner function of the $\Ri$ intermediate states decaying into $D^{*-}\pi^+$. An additional rotation $D^{J_{\pizero(\gamma)}*}_{\lambda'_{\pizero(\gamma)},\lambda_{\pizero(\gamma)}}(\tilde\phi_2,\tilde\theta_2,\tilde\gamma_2)$ is introduced to align the helicity of $\pizero(\gamma)$ with the first decay type. Here, $(\tilde\phi_2,\tilde\theta_2,\tilde\gamma_2)$ represent the Euler angles obtained from Lorentz transform alignment, to match the helicity systems of $\pizero(\gamma)$ in the rest frame of $\dstzero$ of the first decay type~\cite{Wang:2020giv}. This alignment is only applied for a non-zero spin final state. \textcolor{black}{
Details about this rotation can be found in the methodology section {\bf B.4}.
}
This transformation aligns the final state spin in the same helicity coordinate system and allows for the coherent addition of amplitudes for different decay types.

Similarly, in the third subprocess $\ee\to\gammas\to\Ri'\dstminus, \Ri'\to\dstzero\pip,\dstzero\to\dzero\pizero(\gamma), \dstminus\to\dzerobar\pim$, the corresponding decay amplitude is:
\begin{align}\label{app:eq:decay_amp_D3}
&A_{D,3}(\lambda_{\gammas},\lambda_{\pizero(\gamma)})\nonumber\\
=&\sum_{\Ri,\lambda_{\Ri},\lambda'_{\pizero(\gamma)}}A^{\gammas\to \Ri\dstminus}_{\lambda_{\Ri},\lambda_{\dstminus}}\,R(m_{D^{*0}\pi^+})\,A^{\Ri\to\dstzero\pip}_{\lambda_{\dstzero},0}\nonumber\\
&\times A^{\dstminus\to\dzerobar\pim}_{0,0} A^{\dstzero\to\dzero\pizero(\gamma)}_{0,\lambda_{\pizero(\gamma)}}D^{J_{\pizero(\gamma)}*}_{\lambda'_{\pizero(\gamma)},\lambda_{\pizero(\gamma)}}(\tilde\phi_3,\tilde\theta_3,\tilde\gamma_3),
\end{align}

The total amplitude for the process $\ee\to\dstdstpi$ is given by:
\begin{equation}\label{app:}
A_D(\lambda_{\gammas},\lambda_{\pizero(\gamma)})=\sum_{i=1}^{3}c_{i}A_{D,i}(\lambda_{\gammas},\lambda_{\pizero(\gamma)}),
\end{equation}
where $c_{i}$ represents the complex coupling constant that combines with the coupling constant of the reference decay amplitude.

The differential cross section for the process is determined by:
\begin{equation}
\mathrm{d}\sigma\propto\sum_{\lambda_{\gammas},\lambda_{\pizero(\gamma)}}\left|A_D(\lambda_{\gammas},\lambda_{\pizero(\gamma)})\right|^2\mathrm{d}\Phi,
\end{equation}
where $\lambda_{\gammas}=\pm1$, $\lambda_{\pizero}=0$, and $\lambda_{\gamma}=\pm1$. The summation is performed over the helicities $\lambda_{\gamma^*}$ and $\lambda_{\pizero(\gamma)}$. The term $\mathrm{d}\Phi$ represents the element of the standard 3-body phase space.

\subsubsection{2. $\ee\to\pipijpsi$}
The process $\ee\to\pipijpsi$ includes intermediate states $\zc^\pm$ and $\Ri$, represented by
\begin{itemize}
\item[1)] $\ee\to\gammas\to\zcm\pip$, $\zcm\to\pim\jpsi$, $\jpsi\to\lp\lm$,
\item[2)] $\ee\to\gammas\to\zcp\pim$, $\zcp\to\pip\jpsi$, $\jpsi\to\lp\lm$,
\item[3)] $\ee\to\gammas\to\Ri\jpsi$, $\Ri\to\pip\pim$, $\jpsi\to\lp\lm$.
\end{itemize}
To optimise the statistical significance of the spin-parity properties of $\zc$, the PWA fit includes the standard decay of $\jpsi$ into lepton pairs ($\lp\lm$). The helicity amplitude $H^{\jpsi\to \lp\lm}_{\lambda_{\lp},\lambda_{\lm}}$ is set as a constant reference amplitude in the PWA fit.

\begin{figure}[h]
    \includegraphics[width=0.4\textwidth]{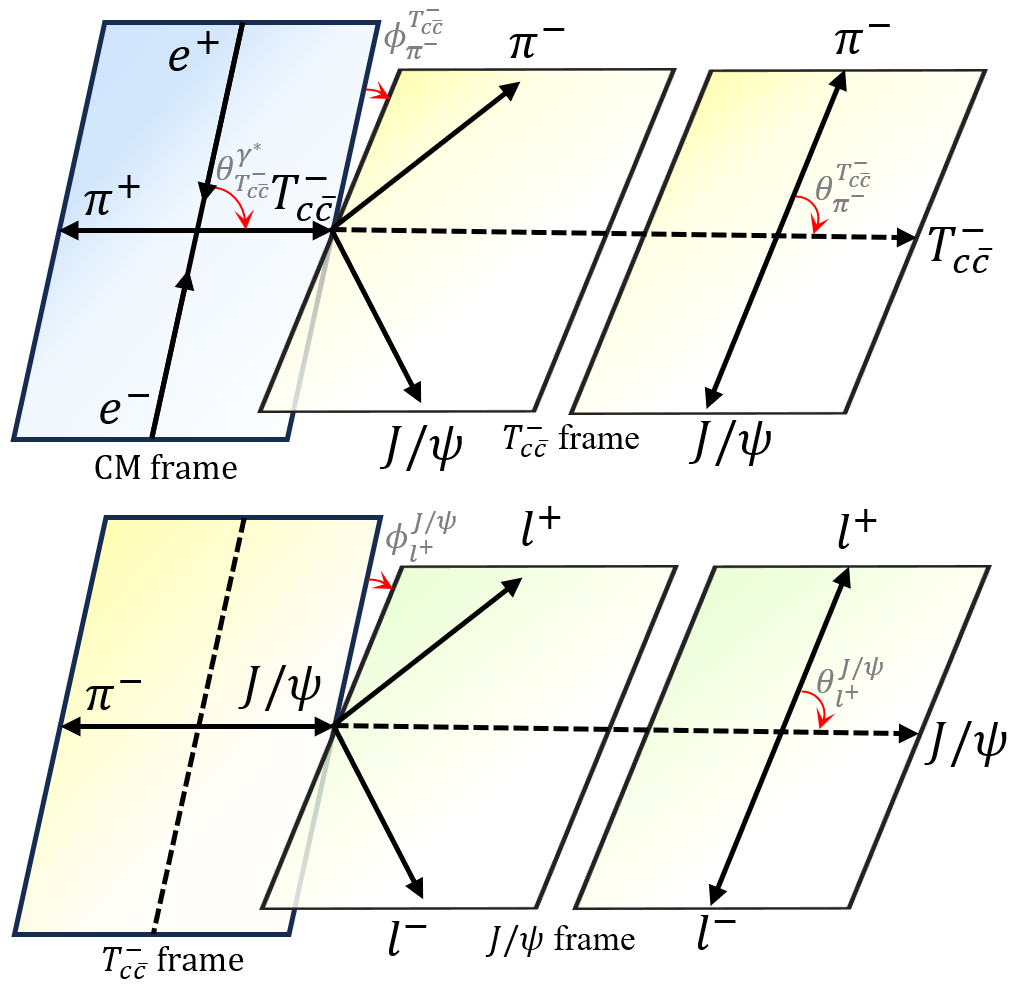}
    \caption{Illustration of the process of $\ee\to\gammas\to\zcm\pip$, $\zcm\to\pim\jpsi$, $\jpsi\to\lp\lm$.}
    \label{app:fig:decay-J-chain1}
\end{figure}

The first subprocess $\ee\to\gammas\to\zcm\pip$, $\zcm\to\pim\jpsi$, $\jpsi\to\lp\lm$, is shown in Fig.~\ref{app:fig:decay-J-chain1}, with the corresponding decay amplitude described by
\begin{align}\label{app:eq:decay_amp_J1}
&A_{J,1}(\lambda_{\gammas},\lambda_{\lp},\lambda_{\lm})\nonumber\\
=&\sum_{\lambda_{\zcm},\lambda_{\jpsi}}A^{\gammas\to\zcm\pip}_{\lambda_{\zcm},0}R(m_{\pi^-\jpsi})A^{\zcm\to\pim\jpsi}_{0,\lambda_{\jpsi}}A^{\jpsi\to\lp\lm}_{\lambda_{\lp},\lambda_{\lm}},
\end{align}
with
\begin{align}
A^{\gammas\to\zcm\pip}_{\lambda_{\zcm},0}&=H^{\gammas\to\zcm\pip}_{\lambda_{\zcm},0}D^{1*}_{\lambda_{\gammas},\lambda_{\zcm}}(\phi_{\zcm}^{\gamma^*},\theta_{\zcm}^{\gamma^*},0),\nonumber\\
A^{\zcm\to\pim\jpsi}_{0,\lambda_{\jpsi}}&=H^{\zcm\to\pim\jpsi}_{0,\lambda_{\jpsi}}D^{J_{\zcm}*}_{\lambda_{\zcm},-\lambda_{\jpsi}}(\phi_{\pim}^{\zcm},\theta_{\pim}^{\zcm},0),\nonumber\\
A^{\jpsi\to\lp\lm}_{\lambda_{\lp},\lambda_{\lm}}&=D^{1*}_{\lambda_{\jpsi},\lambda_{\lp}-\lambda_{\lm}}(\phi_{\lp}^{\jpsi},\theta_{\lp}^{\jpsi},0).\nonumber
\end{align}
Here, we have extracted the Breit-Wigner function for $\jpsi$ and applied it to the MC events. In the $\jpsi$ decay, we have enforced helicity conservation by setting $H^{\jpsi\to\lp\lm}_{1/2,1/2}=H^{\jpsi\to\lp\lm}_{-1/2,-1/2}=0$ and $H^{\jpsi\to\lp\lm}_{-1/2,1/2}=H^{\jpsi\to\lp\lm}_{1/2,-1/2}=1$. The helicity angles are defined similarly to the previous case, as shown in Fig.~\ref{app:fig:decay-J-chain1}.

The amplitude for the second subprocess, $\ee\to\gammas\to\zcp\pim$, $\zcp\to\pip\jpsi$, $\jpsi\to\lp\lm$, is identical to Eq.~\ref{app:eq:decay_amp_J1} but with opposite charges, and is expressed as

\begin{align}
\label{app:eq:decay_amp_J2}
&A_{J,2}(\lambda_{\gammas},\lambda_{\lp},\lambda_{\lm})\nonumber\\
=&\sum_{\lambda_{\zcp},\lambda_{\jpsi},\lambda'_{\lp},\lambda'_{\lm}}A^{\gammas\to\zcp\pim}_{\lambda_{\zcp},0}R(m_{\pi^+\jpsi})A^{\zcp\to\pip\jpsi}_{0,\lambda_{\jpsi}}\nonumber\\
&\times A^{\jpsi\to\lp\lm}_{\lambda'_{\lp},\lambda'_{\lm}}D^{1/2*}_{\lambda'_{\lp},\lambda_{\lp}}(\tilde\phi_2,\tilde\theta_2,\tilde\gamma_2)D^{1/2*}_{\lambda'_{\lm},\lambda_{\lm}}(\tilde\phi'_2,\tilde\theta'_2,\tilde\gamma'_2),
\end{align}
where the term $D^{1/2*}_{\lambda'_{\lp},\lambda_{\lp}}(\tilde\phi_2,\tilde\theta_2,\tilde\gamma_2)D^{1/2*}_{\lambda'_{\lm},\lambda_{\lm}}(\tilde\phi'_2,\tilde\theta'_2,\tilde\gamma'_2)$ is employed to rotate the helicity axis of the leptonic pair by $(\tilde\phi_2,\tilde\theta_2,\tilde\gamma_2)$ and $(\tilde\phi'_2,\tilde\theta'_2,\tilde\gamma'_2)$ in this decay process to align with that in the first decay.

\begin{figure}
\includegraphics[width=0.4\textwidth]{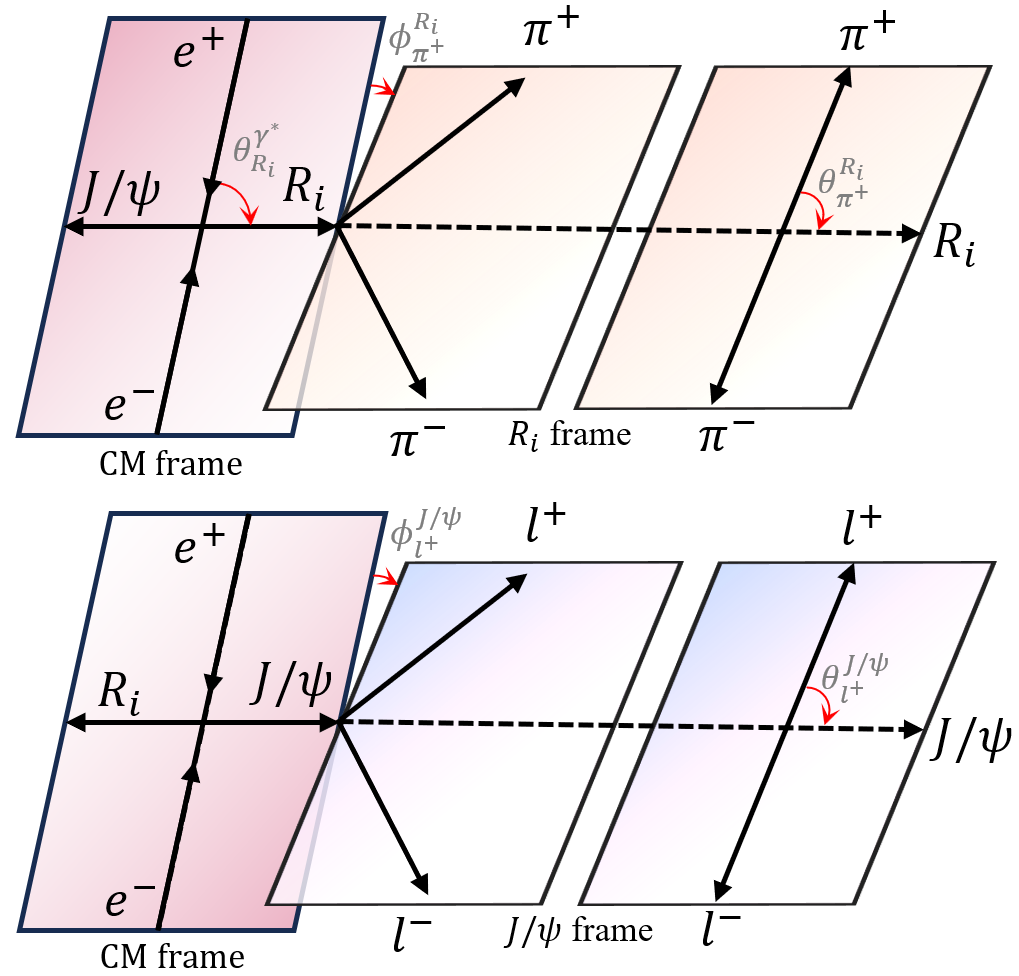}
\caption{Illustration of subprocess of $\ee\to\gammas\to\Ri\jpsi$, $\Ri\to\pip\pim$, $\jpsi\to\lp\lm$.}
\label{app:fig:decay-J-chain3}
\end{figure}

The third subprocess $\ee\to\gammas\to\Ri\jpsi$, $\Ri\to\pip\pim$, $\jpsi\to\lp\lm$, is illustrated in Fig.~\ref{app:fig:decay-J-chain3}, and its decay amplitude is described as
\begin{align}\label{app:eq:decay_amp_J3}
&A_{J,3}(\lambda_{\gammas},\lambda_{\lp},\lambda_{\lm})\nonumber\\
=&\sum_{\Ri,\lambda_{\Ri},\lambda_{\jpsi},\lambda'_{\lp},\lambda'_{\lm}}A^{\gammas\to\Ri\jpsi}_{\lambda_{\jpsi},\lambda_{\Ri}}R_i(m_{\pi^+\pi^-})A^{\Ri\to\pip\pim}_{0,0}\nonumber\\
&\times A^{\jpsi\to\lp\lm}_{\lambda'_{\lp},\lambda'_{\lm}}D^{1/2*}_{\lambda'_{\lp},\lambda_{\lp}}(\tilde\phi_3,\tilde\theta_3,\tilde\gamma_3)D^{1/2*}_{\lambda'_{\lm},\lambda_{\lm}}(\tilde\phi'_3,\tilde\theta'_3,\tilde\gamma'_3),
\end{align}
where the term $R_i$ denotes the Breit-Wigner function for $\Ri$ decaying into $\pipi$, involving intermediate states like $\sigma$, $f_0(980)$, $f_0(1370)$, $f_2(1270)$. Although the nominal masses of $f_0(1370)$ and $f_2(1270)$ are beyond the kinematics of $\gammas\to\pipijpsi$, their widths can influence the low-mass $\pip\pim$ region.

The overall amplitude for $\ee\to\pipijpsi$ is
\begin{equation}\label{app:}
A_J(\lambda_{\gammas},\,\lambda_{\lp},\,\lambda_{\lm})=\sum_{i=1}^\mathrm{3}c_i A_{J,i}(\lambda_{\gammas},\,\lambda_{\lp},\,\lambda_{\lm}),
\end{equation}
where $c_{i}$ represents the complex coupling constant, and the differential cross section is given by
\begin{equation}
\mathrm{d}\sigma\propto\sum_{\lambda_{\gammas},\,\lambda_{\lp},\,\lambda_{\lm}}\left|A_J(\lambda_{\gammas},\,\,\lambda_{\lp},\,\lambda_{\lm})\right|^2\mathrm{d}\Phi,
\end{equation}
where we have $\lambda_{\gammas}=\pm1$, and $\lambda_{\lp},\,\lambda_{\lm}=\pm\frac{1}{2}$ for the final leptonic helicities. The term $\mathrm{d}\Phi$ represents the standard 3-body phase-space element.

\subsubsection{3. $\ee\to\pipihc$}
\label{sec::methods:B3}
The three-body decay is modelled as a two-step process involving sequential two-body decays with an intermediate state, and its spin and parity quantum numbers are assigned according to the intermediate resonance. We take the process $\ee\to\gammas\to\pipihc$ as an example to clarify the amplitude construction.
The full decay amplitude of $\ee\to\pipihc$ consists of three subprocesses:
\begin{itemize}
\item[1)] $\ee\to\gammas\to\zcm\pip$, $\zcm\to\pim\hc$, $\hc\to\gamma\etac$,
\item[2)] $\ee\to\gammas\to\zcp\pim$, $\zcp\to\pip\hc$, $\hc\to\gamma\etac$,
\item[3)] $\ee\to\gammas\to R_{\pip\pim}\hc$, $R_{\pip\pim}\to\pip\pim$, $\hc\to\gamma\etac$.
\end{itemize}
Here, the $\zc^{\mp}$ denotes the resonances on $\pi^{\mp}\hc$ and the $R_{\pip\pim}$ for $\pip\pim$. The common decay $\hc\to\gamma\etac$ is included in the PWA to optimise the statistical significance of the spin-parity properties of the $\zc(4020)^{\pm}$. The helicity amplitude $H^{\hc\to\gamma\etac}_{\lambda_{\gamma},0}$ is used as the reference amplitude by setting it to a constant value in the PWA fit.

For $\ee\to\gammas\to\zcm\pip$, $\zcm\to\pim\hc$,
\begin{figure}[h]
    \includegraphics[width=0.4\textwidth]{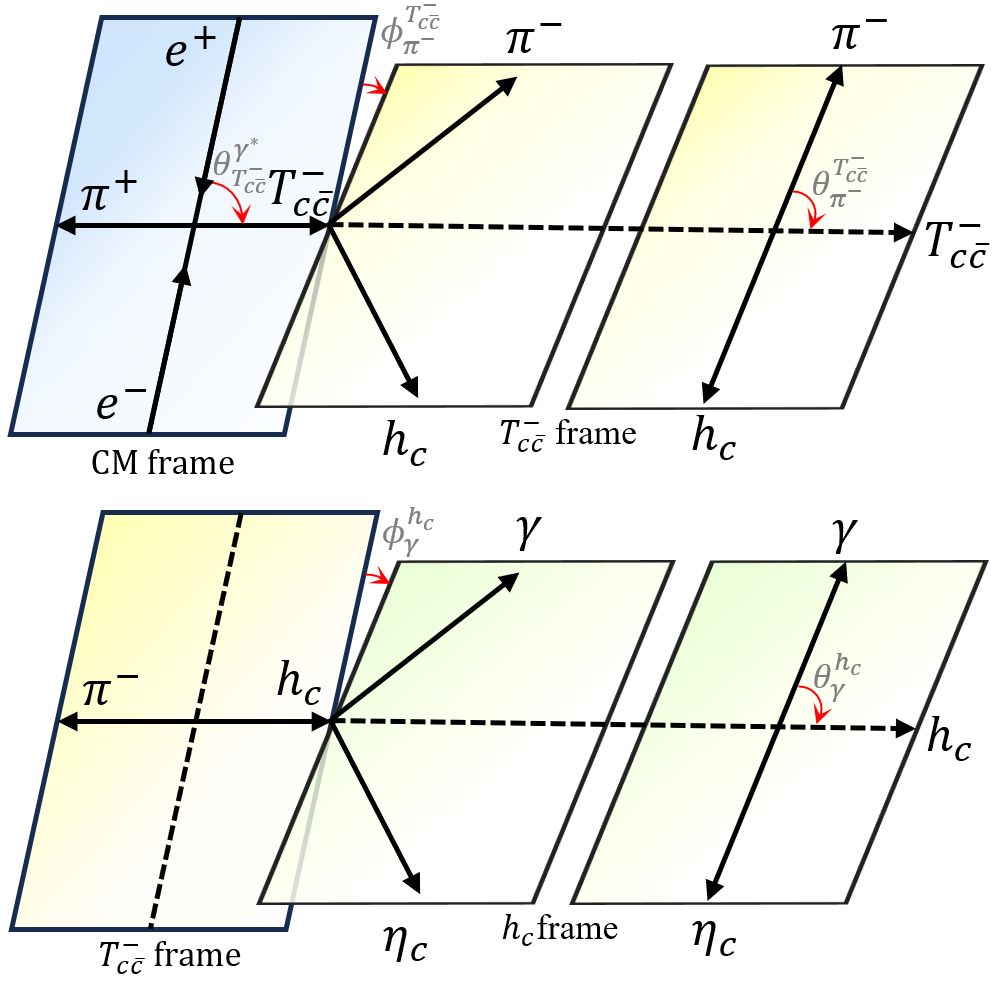}
    \caption{An illustration and definition of the various angles related to the subprocess of $\ee\to\gammas\to\zcm\pip$, $\zcm\to\pim\hc$, $\hc\to\gamma\etac$.}
    \label{fig:decay-H-chain1}
\end{figure}
the amplitude for the first subprocess $\ee\to\pipihc$ with $\zc^-$ state can be written as
\begin{align}\label{eq:decay_amp_H1}
&A_{H,1}(\lambda_{\gammas},\,\lambda_{\gamma})\nonumber\\
=&\sum_{\lambda_{\zcm},\,\lambda_{\hc}}A^{\gammas\to\zcm\pip}_{\lambda_{\zcm},\,0}R(m_{\pi^-h_c})A^{\zcm\to\pim\hc}_{0,\,\lambda_{\hc}}A^{\hc\to\gamma\etac}_{\lambda_{\gamma},\,0},
\end{align}
with
\begin{align}
A^{\gammas\to\zcm\pip}_{\lambda_{\zcm},\,0}&=H^{\gammas\to\zcm\pip}_{\lambda_{\zcm},\,0}D^{1*}_{\lambda_{\gammas},\,\lambda_{\zcm}}(\phi_{\zcm}^{\gamma^*},\,\theta_{\zcm}^{\gamma^*},\,0),\nonumber\\
A^{\zcm\to\pim\hc}_{0,\,\lambda_{\hc}}&=H^{\zcm\to\pim\hc}_{0,\,\lambda_{\hc}}D^{J_{\zcm}*}_{\lambda_{\zcm},\,-\lambda_{\hc}}(\phi_{\pim}^{\zcm},\,\theta_{\pim}^{\zcm},\,0),\nonumber\\
A^{\hc\to\gamma\etac}_{\lambda_{\gamma},\,0}&=D^{1*}_{\lambda_{\hc},\,\lambda_{\gamma}}(\phi_{\gamma}^{\hc},\,\theta_{\gamma}^{\hc},\,0),\nonumber
\end{align}
where all the definition is following the mentioned cases.

The amplitude of the charge conjugated decay of the $\zc^+$ state, $\ie$, $\ee\to\gammas\to\zcp\pim, \zcp\to\pip\hc$, $\hc\to\gamma\etac$, is similar to Eq.~\ref{eq:decay_amp_H1} and can be expressed as

\begin{align}\label{eq:decay_amp_H2}
&A_{H,2}(\lambda_{\gammas},\lambda_{\gamma})\nonumber\\
=&\sum_{\lambda_{\zcp},\,\lambda_{\hc},\,\lambda_{\gamma'}}A^{\gammas\to\zcp\pim}_{\lambda_{\zcp},\,0}R(m_{\pi^+h_c})A^{\zcp\to\pip\hc}_{0,\,\lambda_{\hc}}A^{\hc\to\gamma\etac}_{\lambda_{\gamma}',\,0}\nonumber\\
&\times D^{1*}_{\lambda_{\gamma}',\,\lambda_{\gamma}}(\tilde{\phi}_{2},\,\tilde{\theta}_{2},\,\tilde{\gamma}_{2}).
\end{align}
An additional rotation $D^{1*}_{\lambda_{\gamma}',\,\lambda_{\gamma}}(\tilde{\phi}_{2},\,\tilde{\theta}_{2},\,\tilde{\gamma}_{2})$ is introduced to align the photon helicity with the first decay type, where $(\tilde{\phi}_{2},\,\tilde{\theta}_{2},\,\tilde{\gamma}_{2})$ is the angle by which the photon helicity axis is rotated to match that of the first subprocess.

\begin{figure}[h]
    \includegraphics[width=0.4\textwidth]{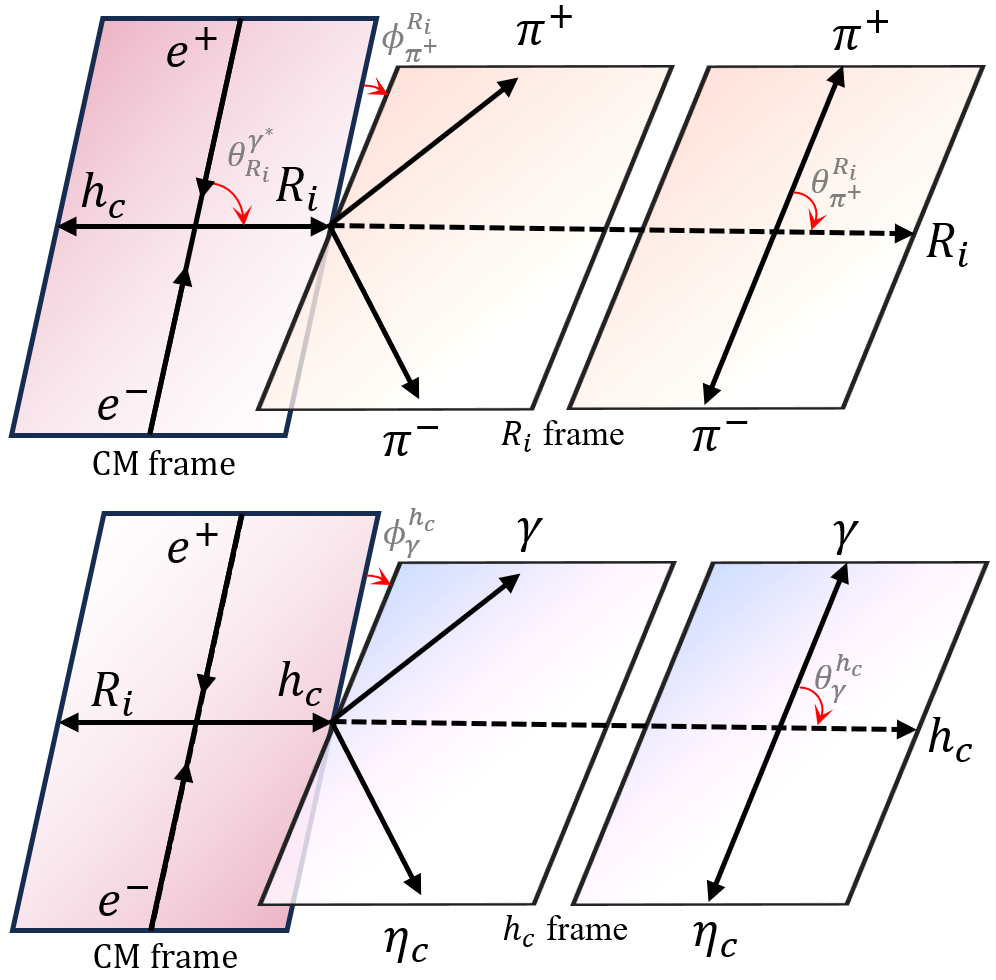}
    \caption{An illustration of the subprocess of $\ee\to\gammas\to\Ri\hc$, $\Ri\to\pip\pim$, $\hc\to\gamma\etac$, defining the angles as used in our formalism.}
    \label{fig:decay-H-chain3}
\end{figure}

The third subprocess $\ee\to\gammas\to\Ri\hc$, $\Ri\to\pip\pim$, $\hc\to\gamma\etac$, is sketched in Fig.~\ref{fig:decay-H-chain3}, and the amplitude reads
\begin{align}\label{eq:decay_amp_H3}
&A_{H,\,3}(\lambda_{\gammas},\,\lambda_{\gamma})\nonumber\\
=&\sum_{\Ri,\,\lambda_{\Ri},\,\lambda_{\hc},\,\lambda_{\gamma'}}A^{\gammas\to\hc\Ri}_{\lambda_{\hc},\,\lambda_{\Ri}}R(m_{\pip\pim})A^{\Ri\to\pip\pim}_{0,\,0}A^{\hc\to\gamma\etac}_{0,\,\lambda'_{\gamma}}\nonumber\\
&\times D^{1*}_{\lambda_{\gamma}',\,\lambda_{\gamma}}(\tilde\phi_3,\,\tilde\theta_3,\,\tilde\gamma_3),
\end{align}
with
\begin{align}
A^{\gammas\to\hc\Ri}_{\lambda_{\hc},\,\lambda_{\Ri}}&=H^{\gammas\to\hc\Ri}_{\lambda_{\hc},\,\lambda_{\Ri}}D^{1*}_{\lambda_{\gammas},\,\lambda_{\hc}-\lambda_{\Ri}}(\phi_{\hc}^{\gamma^*},\,\theta_{\hc}^{\gamma^*},\,0),\nonumber\\
A^{\Ri\to\pip\pim}_{0,\,0}&=H^{\Ri\to\pip\pim}_{0,\,0}D^{J_{\Ri}*}_{\lambda_{\Ri},\,0}(\phi_{\pip}^{\Ri},\,\theta_{\pip}^{\Ri},\,0),\nonumber\\
A^{\hc\to\gamma\etac}_{\lambda'_{\gamma},\,0}&=D^{1*}_{\lambda_{\hc},\,\lambda'_{\gamma}}(\phi_{\gamma}^{\hc},\,\theta_{\gamma}^{\hc},\,0),\nonumber
\end{align}
where $\Ri$ is a resonance that decays into $\pipi$, such as $\sigma$ or $f_0(980)$ intermediate states. The central value of the $f_0(980)$ mass is outside the kinematics of $\gammas\to\pipihc$. However, given the width of this state, it may kinematically impact the high $\pipi$ mass region. The helicity angles in Eq.~\eqref{eq:decay_amp_H3} have similar definitions, and the additional rotation $D^{1*}_{\lambda_{\gamma}',\,\lambda_{\gamma}}(\tilde\phi_3,\,\tilde\theta_3,\,\tilde\gamma_3)$ also has the same purpose as mentioned previously.

The total amplitude for $\ee\to\pipihc$ is expressed by:
\begin{equation}
A_H(\lambda_{\gammas},\,\lambda_{\gamma})=\sum_{i=1}^\mathrm{3}c_i A_{H,\,i}(\lambda_{\gammas},\,\lambda_{\gamma}),
\end{equation}
where $c_{i}$ is the complex coupling constant, which recombines the coupling constant of the reference decay amplitude.

The partial decay rate of $\gamma^*$ is given by:
\begin{equation}
\mathrm{d}\sigma\propto\sum_{\lambda_{\gammas},\,\lambda_{\gamma}}\left|A_H(\lambda_{\gammas},\,\lambda_{\gamma})\right|^2\mathrm{d}\Phi,
\end{equation}
where $\lambda_{\gammas}=\pm1$ due to the helicity conservation in the $\gammas$ coupling to $\ee$, and $\lambda_{\gamma}=\pm1$ for the radiative photon in the final state. The parameter $\mathrm{d}\Phi$ represents the element of the three-body PHSP.

\subsubsection{4. Alignment rotation}
Under the $SU(2)$ group, the rotation and boost operations can be expressed as
\begin{align*}
R_{z}(\phi) &=
\begin{pmatrix}
e^{-i\frac{\phi}{2}} & 0 \\
0 & e^{i\frac{\phi}{2}}
\end{pmatrix},\,
R_{y}(\theta) =
\begin{pmatrix}
\cos\frac{\theta}{2} & -\sin\frac{\theta}{2} \\
\sin\frac{\theta}{2} &  \cos\frac{\theta}{2}
\end{pmatrix}, \\
B_{z}(\omega) &=
\begin{pmatrix}
e^{-\frac{\omega}{2}} & 0 \\
0 & e^{\frac{\omega}{2}}
\end{pmatrix},
\end{align*}
where $\omega=\tanh^{-1}{\frac{|\bf{p}|}{E}}$, with $\bf{p}$ and ${E}$ the boosted momentum and energy, respectively. $\theta$ and $\phi$ are the helicity angles of the decay. The alignment rotation can be expressed by a 2-D matrix. For instance, in the first (reference) subprocess $\ee\to\gammas\to\zcm\pip$, $\zcm\to\pim\hc$, $\hc\to\gamma\etac$, the total Lorentz transform of $\gamma$ is
\begin{align}
L_{\zcm}=&R_{y}(\theta^{\ee}_{\zcm})R_{z}(\phi^{\ee}_{\zcm})B_{z}(\omega^{\ee}_{\zcm})R_{y}(\theta^{\zcm}_{\hc})R_{z}(\phi^{\zcm}_{\hc})\nonumber\\
&\cdot B_{z}(\omega^{\zcm}_{\hc})R_{y}(\theta^{\hc}_{\gamma})R_{z}(\phi^{\hc}_{\gamma})B_{z}(\omega^{\hc}_{\gamma}).
\end{align}
Similar in second subprocess $\ee\to\gammas\to\zcp\pim$, $\zcp\to\pip\hc$, $\hc\to\gamma\etac$, the corresponding Lorentz transform of $\gamma$ is
\begin{align}
L_{\zcp}=&R_{y}(\theta^{\ee}_{\zcp})R_{z}(\phi^{\ee}_{\zcp})B_{z}(\omega^{\ee}_{\zcp})R_{y}(\theta^{\zcp}_{\hc})R_{z}(\phi^{\zcp}_{\hc})\nonumber\\
&\cdot B_{z}(\omega^{\zcp}_{\hc})R_{y}(\theta'^{\hc}_{\gamma})R_{z}(\phi'^{\hc}_{\gamma})B_{z}(\omega'^{\hc}_{\gamma}).
\end{align}
The relative Lorentz transform between the two coordinate systems defined for $h_c\to \gamma \eta_c$ decays is
\begin{equation}
L_{\rm align}(\tilde{\phi}_{2},\,\tilde{\theta}_{2},\,\tilde{\gamma}_{2}) R_{y}(\theta'^{\hc}_{\gamma})R_{z}(\phi'^{\hc}_{\gamma}) =R_{y}(\theta^{\hc}_{\gamma})R_{z}(\phi^{\hc}_{\gamma}),
\end{equation}
and the final Euler angle $(\tilde{\phi}_{2},\,\tilde{\theta}_{2},\,\tilde{\gamma}_{2})$ can be determined from it. This transformation aligns the helicities of the final particles within a common helicity coordinate system, enabling the coherent addition of amplitudes from different decay types.

\subsubsection{C. Likelihood Function in PWA Fit}
The complex coupling constants $g_{LS}$, the resonance mass of $\zc(4020)^{-}$ and the value of $\overline{\sum}|H_{X'}|^2$ are determined by an unbinned maximum-likelihood fit to data.
The likelihood function of each channel at one energy point is constructed according to~\cite{Langenbruch:2019nwe}:
\begin{equation}
\label{eq:llh}
-\ln\mathcal{L} = -\alpha\left(\sum_{i=1}^{N_\mathrm{data}} \ln p^{(i)}_{\mathrm{data}}-\mathcal{W}_\mathrm{bkg}\sum_{i=1}^{N_{\mathrm{bkg}}}\ln p^{(i)}_{\mathrm{bkg}}-\mu\right),
\end{equation}
where $\alpha=\frac{N_\mathrm{data}-\mathcal{W}_\mathrm{bkg}N_\mathrm{bkg}}{N_\mathrm{data}+\mathcal{W}^2_\mathrm{bkg}N_\mathrm{bkg}}$ is a normalization factor to account for the background level~\cite{Langenbruch:2019nwe} and the numbers are quoted from Table~\ref{tab:dataset}., $p^{(i)}_{\mathrm{data}(\mathrm{bkg})}=\epsilon(x_i)|A(x_{i})|^{2}$ is the likelihood function for $i$-th data (background) event, while $\mu=\int\epsilon(x_i)|A(x_{i})|^{2}\mathrm{d}\Phi$ represents the likelihood value from integral MC sample.
The function $\epsilon(x_i)$ is the detection efficiency, which depends on the four-momentum $x_{i}$ of final particles. 
$\mathcal{W}_\mathrm{bkg}$ is the weight factor for each background event, 
and is given by the ratio of the lengths of the signal and sideband ranges.
The modulus squared $|A|^{2}$ for the three channels is defined as
\[
|A|^{2} =
\left\{
\begin{aligned}
& \sum_{\lambda_{\gammas},\,\lambda_{\pizero(\gamma)}}\left|A_{D}(\lambda_{\gammas},\,\lambda_{\pizero(\gamma)})\right|^2,\\
& \sum_{\lambda_{\gammas},\,\lambda_{\lp},\,\lambda_{\lm}}\left|A_J(\lambda_{\gammas},\,\lambda_{\lp},\,\lambda_{\lm})\right|^2, \\
&
\sum_{\lambda_{\gammas},\,\lambda_{\gamma}}\left|A_H(\lambda_{\gammas},\,\lambda_{\gamma})\right|^2.
\end{aligned}
\right.
\]
To perform a simultaneous fit to data corresponding to the three signal processes at the two energy points, the likelihood $\ln\mathcal{L}$ is independently calculated for each data set and then aggregated to form the objective function across the six data sets as
\begin{eqnarray}
\label{eq:S}
S=-\ln\mathcal{L}=-\sum_{n=1}^{6} \ln \mathcal{L}_n,
\end{eqnarray}
which is minimized with {\sc minuit}~\cite{James:1975dr}.
The leading-order rates of these three-body decay processes are incorporated to further constrain the coupling parameters used in extracting $\Gamma(m)$. In the PWA fit, the integral of each MC sample is fixed to the leading-order rates for these three processes.

\subsubsection{D. Full Pole Positions}
The final states from three channels are separated, so the pole singularities can be calculated to locate the pole positions on the complex momentum plane. The three branch points at each channel threshold generate an eight-sheet Riemann surface, and the momentum signs are assigned accordingly. The full pole positions are summarized in Table~\ref{tab:S1_pole}.
\begin{table}[htbp]
	\caption{Summary of pole positions of $\zc(4020)^{-}$. The uncertainties are statistical only and propagated through the corresponding covariance matrix.
    The Imag.($p$) represent the signs of imaginary part of momenta of $\jpsi\pim$, $\hc\pim$, and $\dstzero\dstminus$ channels, respectively. The symbol `$+$' (`$-$') means the imaginary part of the corresponding momentum is positive (negative).
    }
	\label{tab:S1_pole}
	\begin{center}
		\begin{tabular}{lccc}
			\hline\hline
			Pole & $M_{\mathrm{pole}}$ ($\mevcc$) & $\Gamma_{\mathrm{pole}}$ ($\mev$) & Imag.($p$)  \\\hline
			1    & $4022.44\pm1.55$ &  $38.54\pm2.94$  & $(-,\,-,\,-)$ \\
			2    & $4023.01\pm1.35$ &  $35.02\pm2.20$  & $(-,\,+,\,-)$ \\
			3    & $4029.39\pm0.86$ &   $0.03\pm15.66$ & $(-,\,+,\,+)$ \\
			4    & $4029.87\pm0.95$ &   $2.85\pm14.56$ & $(-,\,-,\,+)$ \\
			5    & $4022.44\pm1.55$ & $-38.54\pm2.94$  & $(-,\,-,\,-)$ \\
			6    & $4023.01\pm1.35$ & $-35.02\pm2.20$  & $(-,\,+,\,-)$ \\
			7    & $4029.39\pm0.86$ &  $-0.03\pm15.66$ & $(-,\,+,\,+)$ \\
			8    & $4029.87\pm0.95$ &  $-2.85\pm14.56$ & $(-,\,-,\,+)$ \\
			\hline \hline
		\end{tabular}
    \end{center}
\end{table}

\begin{table*}[htbp]
	\begin{center}
    \caption{Summary of the $\jp$ hypothesis test of the $\zc(4020)^{-}$ and the statistical significance of the $1^+$  over other spin-parity assignments.}
	\label{tab:S2_jp}
	\begin{tabular}{c c c c c}
		\hline\hline
		$H_0$ & $H_1$   & $\Delta(-2\ln \mathcal{L})$ & $\Delta(\mathrm{ndof})$ & Statistical significance\\
		\hline
		only $1^-$ & $1^-$ with additional $1^+$ & 409.2  & 23 & $18.0\sigma$\\
		only $2^+$ & $2^+$ with additional $1^+$ & 435.4  & 23 & $18.7\sigma$\\
		only $2^-$ & $2^-$ with additional $1^+$ & 329.4  & 23 & $15.8\sigma$\\
		\hline
		only $1^+$ & $1^+$ with additional $1^-$  & 27.8 & 21 & $1.5\sigma$\\
		only $1^+$ & $1^+$ with additional $2^+$  & 30.2 & 23 & $1.5\sigma$\\
		only $1^+$ & $1^+$ with additional $2^-$  & 56.4 & 25 & $3.6\sigma$\\
		\hline\hline
	\end{tabular}
	\end{center}
\end{table*}
\subsubsection{E. Hypothesis test of spin-parity in data}
\textcolor{black}{
Table~\ref{tab:S2_jp} shows the significance when attributing $1^+$ to the $\zc(4020)^{-}$ compared to other spin-parity assignments. Here, $\Delta(\mathrm{ndof})=23$ encompasses the parameters of the $1^+$ $\zc(4020)^{-}$ resonance, which include coupling constants and multi-channel BW parameters in the PWA fit to two data sets at $\sqrt{s}=4.395$ and $4.416\gev$.
In all scenarios, the significance of the hypothesis $1^+$ exceeds $15.8\sigma$ in comparison to alternative hypotheses. In contrast, the significance of attributing a different spin-parity to the $\zc(4020)^{-}$ over $1^+$ is less than $3.6\sigma$.}

%% ends here %%

%%%%%%%%%%%%%%%%%%%%%%%%%%%%%%%%%%%%%%%%%%%%%%%%%%%%%%%%%%%%%%%%
%%%%%    bibliographies       Part                %%%%%%%%%%%%%
%%%%%%%%%%%%%%%%%%%%%%%%%%%%%%%%%%%%%%%%%%%%%%%%%%%%%%%%%%%%%%%%
\clearpage
\bibliography{reference}%

\end{document}

%% file: authorlist_2025-03-21.tex
%% Saved at => 2025-03-21
M.~Ablikim$^{1}$, M.~N.~Achasov$^{4,c}$, P.~Adlarson$^{77}$, X.~C.~Ai$^{82}$, R.~Aliberti$^{36}$, A.~Amoroso$^{76A,76C}$, Q.~An$^{73,59,a}$, Y.~Bai$^{58}$, O.~Bakina$^{37}$, Y.~Ban$^{47,h}$
% ...
\\
\vspace{0.2cm}
(BESIII Collaboration)\\
\vspace{0.2cm} {\it
$^{1}$ Institute of High Energy Physics, Beijing 100049, People's Republic of China\\
% ...
$^{4}$ Budker Institute of Nuclear Physics SB RAS (BINP), Novosibirsk 630090, Russia\\
% ...
$^{36}$ Johannes Gutenberg University of Mainz, Johann-Joachim-Becher-Weg 45, D-55099 Mainz, Germany\\
$^{37}$ Joint Institute for Nuclear Research, 141980 Dubna, Moscow region, Russia\\
% ...
$^{47}$ Peking University, Beijing 100871, People's Republic of China\\
% ...
$^{58}$ Southeast University, Nanjing 211100, People's Republic of China\\
$^{59}$ State Key Laboratory of Particle Detection and Electronics, Beijing 100049, Hefei 230026, People's Republic of China\\
% ...
$^{73}$ University of Science and Technology of China, Hefei 230026, People's Republic of China\\
% ...
$^{76}$ University of Turin and INFN, (A)University of Turin, I-10125, Turin, Italy; (B)University of Eastern Piedmont, I-15121, Alessandria, Italy; (C)INFN, I-10125, Turin, Italy\\
$^{77}$ Uppsala University, Box 516, SE-75120 Uppsala, Sweden\\
% ...
$^{82}$ Zhengzhou University, Zhengzhou 450001, People's Republic of China\\

\vspace{0.2cm}
$^{a}$ Deceased\\
% ...
$^{c}$ Also at the Novosibirsk State University, Novosibirsk, 630090, Russia\\
% ...
$^{h}$ Also at State Key Laboratory of Nuclear Physics and Technology, Peking University, Beijing 100871, People's Republic of China\\
% ...

}
%% ends here %%

%% file: acknowledgement_2025-03-21.tex
%% Saved at => 2025-03-21
%\textbf{Acknowledgement}

The BESIII Collaboration thanks the staff of BEPCII and the IHEP computing center for their strong support. This work is supported in part by National Key R\&D Program of China under Contracts Nos. 2020YFA0406300, 2020YFA0406400, 2023YFA1606000, 2023YFA1606704; National Natural Science Foundation of China (NSFC) under Contracts Nos. 11635010, 11935015, 11935016, 11935018, 12025502, 12035009, 12035013, 12061131003, 12175244, 12192260, 12192261, 12192262, 12192263, 12192264, 12192265, 12221005, 12225509, 12235017, 12361141819, 12422504; the Chinese Academy of Sciences (CAS) Large-Scale Scientific Facility Program; CAS under Contract No. YSBR-101; 100 Talents Program of CAS; Fundamental Research Funds for the Central Universities, Lanzhou University, University of
Chinese Academy of Sciences; The Institute of Nuclear and Particle Physics (INPAC) and Shanghai Key Laboratory for Particle Physics and Cosmology; 
%Agencia Nacional de Investigación y Desarrollo de Chile (ANID), Chile under Contract No. ANID PIA/APOYO AFB230003; 
German Research Foundation DFG under Contract No. FOR5327; Istituto Nazionale di Fisica Nucleare, Italy; Knut and Alice Wallenberg Foundation under Contracts Nos. 2021.0174, 2021.0299; Ministry of Development of Turkey under Contract No. DPT2006K-120470; National Research Foundation of Korea under Contract No. NRF-2022R1A2C1092335; National Science and Technology fund of Mongolia; Polish National Science Centre under Contract No. 2024/53/B/ST2/00975; Swedish Research Council under Contract No. 2019.04595; U. S. Department of Energy under Contract No. DE-FG02-05ER41374.

%% ends here %%